\documentclass[fleqn,usenatbib]{mnras}

\usepackage[T1]{fontenc}

\DeclareRobustCommand{\VAN}[3]{#2}
\let\VANthebibliography\thebibliography
\def\thebibliography{\DeclareRobustCommand{\VAN}[3]{##3}\VANthebibliography}

\DeclareRobustCommand{\DE}[3]{#2}
\let\DEthebibliography\thebibliography
\def\thebibliography{\DeclareRobustCommand{\DE}[3]{##3}\DEthebibliography}

\usepackage{graphicx}	
\usepackage{amsmath}	
\usepackage{amssymb}	
\usepackage{siunitx}    
\usepackage{bm}
\usepackage{newtxtext,newtxmath}

\newcommand{\logNHI}{\ensuremath{\log N(\mbox{\ion{H}{i}})}}
\newcommand{\logNHIunit}{\ensuremath{\log [N(\mbox{\ion{H}{i}})/\rm{cm}^{-2}]}}
\newcommand{\kms}{\ensuremath{\text{km s}^{-1}} }
\newcommand{\Moyr}{\ensuremath{\text{M}_{\odot}\ \text{yr}^{-1}}}
\DeclareUnicodeCharacter{0308}{}
\newcommand\bolden[1]{{\boldmath\bfseries#1}}

\title[Statistical Study of CGM Gas]{MUSE-ALMA Haloes VIII: Statistical Study of Circumgalactic Medium Gas}

\author[Weng et al.]{
S. Weng,$^{1,2,3,4}$\thanks{E-mail: simon.weng@eso.org}
{C. P\'eroux},$^{1,5}$
A. Karki,$^{6}$ 
R. Augustin,$^{7}$ 
V. P. Kulkarni,$^{6}$ 
R. Szakacs,$^{1}$
\newauthor
M. A. Zwaan,$^{1}$
A. Klitsch,$^{8}$ 
A. Hamanowicz,$^{7}$
E. M. Sadler,$^{2, 3, 4}$
A. Biggs,$^{1}$
A. Y. Fresco,$^{9}$ 
\newauthor
M. Hayes,$^{10}$ 
J. C. Howk,$^{11}$
G. G. Kacprzak,$^{12, 3}$
H. Kuntschner,$^{1}$
D. Nelson,$^{13}$ 
\& M. Pettini$^{14}$
\\
$^{1}$ European Southern Observatory, Karl-Schwarzschildstrasse 2, D-85748 Garching bei M{\"u}nchen, Germany\\
$^2$ Sydney Institute for Astronomy, School of Physics A28, University of Sydney, NSW 2006, Australia\\
$^3$ ARC Centre of Excellence for All Sky Astrophysics in 3 Dimensions (ASTRO 3D)\\ 
$^4$ ATNF, CSIRO Space and Astronomy,  PO Box 76, Epping, NSW 1710, Australia \\
$^{5}$ Aix Marseille Universit\'e, CNRS, LAM (Laboratoire d'Astrophysique de Marseille) UMR 7326, 13388, Marseille, France \\
$^6$ Department of Physics and Astronomy, University of South Carolina, Columbia, SC 29208, USA\\
$^{7}$ Space Telescope Science Institute, 3700 San Martin Drive, Baltimore, MD 21218, USA \\
$^{8}$ DARK, Niels Bohr Institute, University of Copenhagen, Jagtvej 128, 2200 Copenhagen, Denmark\\
$^{9}$ Max-Planck-Institut f{\"u}r Extraterrestrische Physik (MPE), Giessenbachstrasse 1, D--85748 Garching, Germany\\
$^{10}$ Stockholm University, Department of Astronomy and Oskar Klein Centre for Cosmoparticle Physics, AlbaNova University Centre, SE-10691, Stockholm, Sweden\\
$^{11}$ Department of Physics, University of Notre Dame, Notre Dame, Indiana 46556, USA\\
$^{12}$ Centre for Astrophysics and Supercomputing, Swinburne University of Technology, Hawthorn, Victoria 3122, Australia\\
$^{13}$ Universit\"at Heidelberg, Zentrum f\"ür Astronomie, Institut f\"ur theoretische Astrophysik, Albert-Ueberle-Str. 2, 69120 Heidelberg, Germany\\
$^{14}$ Institute of Astronomy, University of Cambridge, Madingley Road, Cambridge CB3 0HA, UK\\
}

\date{Accepted XXX. Received YYY; in original form ZZZ}

\pubyear{2022}

\begin{document}
\label{firstpage}
\pagerange{\pageref{firstpage}--\pageref{lastpage}}
\maketitle

\begin{abstract}
The distribution of gas and metals in the circumgalactic medium (CGM) plays a critical role in how galaxies evolve. 
The MUSE-ALMA Halos survey combines MUSE, ALMA and HST observations to constrain the properties of the multi-phase gas in the CGM and the galaxies associated with the gas probed in absorption.
In this paper, we analyse the properties of galaxies associated with 32 strong \ion{H}{i} Ly-$\alpha$ absorbers at redshift $0.2 \lesssim z \lesssim 1.4$.
We detect 79 galaxies within $\pm 500$ \kms \!of the absorbers in our 19 MUSE fields.
These associated galaxies are found at physical distances from 5.7 kpc and reach star-formation rates as low as $0.1$ \Moyr. 
The significant number of associated galaxies allows us to map their physical distribution on the $\Delta v$ and $b$ plane.
Building on previous studies, we examine the physical and nebular properties of these associated galaxies and find the following: i) 27/32 absorbers have galaxy counterparts and more than 50 per cent of the absorbers have two or more associated galaxies, ii) the \ion{H}{i} column density of absorbers is anti-correlated with the impact parameter (scaled by virial radius) of the nearest galaxy as expected from simulations, iii) the metallicity of associated galaxies is typically larger than the absorber metallicity which decreases at larger impact parameters.
It becomes clear that while strong \ion{H}{i} absorbers are typically associated with more than a single galaxy, we can use them to statistically map the gas and metal distribution in the CGM.
\end{abstract}



\section{Introduction}
The circumgalactic medium (CGM) extends beyond a galaxy's disk and interstellar medium (ISM) to the intergalactic medium (IGM) and is the region where inflowing, outflowing and recycled gas transits \citep{Tumlinson2017}.
Inflowing metal-poor gas from dark matter filaments is expected to fuel star formation within galaxies, while galactic winds eject baryons into the halo or even the IGM at larger velocities.
A portion of this wind material is expected to be recycled onto the galaxy in the form of galactic fountains \citep{Fraternali2017}.
The combination of these processes results in a continuous cycling of baryons within and outside galaxies, and dictates how galaxies evolve \citep{PerouxHowk2020}. 
Observationally linking the distribution of baryons in the circumgalactic medium with galaxy properties is critical to constraining the role of the CGM in galaxy evolution.

Characterizing the distribution of matter in the CGM beyond the local Universe has been challenging because of the diffuse nature of material that comprises it. 
Despite the success of several studies probing the CGM in emission, the systems discovered so far contain powerful active galactic nuclei (AGN), starbursting galaxies or are highly overdense regions \citep[e.g.][]{Epinat2018, Johnson2018, Chen2019, Helton2021, Burchett2021, Cameron2021}. 
Studies of systems more representative of `average' galaxies and galaxy groups often require significant exposure times to obtain a detection in emission \citep{Zabl2021, Leclercq2022} or stacking \citep{Steidel2011, Momose2014, Chen2020b}. 
While emission-line studies enable the spatial mapping of the CGM, absorption-line spectroscopy allows the CGM gas to be detected to much lower column densities. 
By observing the single sightlines towards bright background quasi-stellar objects (QSOs), the column densities probed are only limited by the apparent brightness of the background source and reach far greater depths than what can be observed in emission. 
To remedy the fact that this technique only probes a single sightline through a galaxy’s CGM, large samples are required to statistically map the distribution of baryons and place observational constraints that can inform simulations. 

For over two decades, the identification of galaxies associated with absorbers in QSO spectra \citep[e.g.][]{Bergeron1986, BergeronBoisse1991} required both imaging and spectroscopy.
This was often inefficient as the targets selected for spectroscopic follow-up due to their proximity to the QSO were found to be not always at the absorber redshift.
The advent of integral field spectroscopy (IFS) meant that associated galaxies could now be discovered in a more unbiased and efficient manner, and the sample of quasar-galaxy pairs has since increased in size by more than two orders of magnitude.
In particular, the Multi-Unit Spectrographic Explorer (MUSE) instrument \citep{Bacon2010}, with its wide $1 \times 1$ arcmin field-of-view (FoV) and simultaneous wavelength coverage $4700 - 9350$ \AA, has led to a proliferation of surveys characterizing the CGM by studying galaxies in the vicinity of absorbers.
The programs MusE GAs FLOw and Wind \citep[MEGAFLOW][]{Bouche2016} and MUSE Analysis of Gas around Galaxies \citep[MAGG][]{Lofthouse2020} span redshift $0.2 \lesssim z \lesssim 4$ and probe gas in the CGM through various tracers such as \ion{Mg}{II} ($\lambdaup \lambdaup 2795, 2803$) and \ion{H}{i} absorption.
The MUSE Quasar-field Blind Emitters Survey \citep[MUSEQuBES][]{Muzahid2020} target bright high-$z$ QSOs with absorption lines to search for Lyman-$\alpha$ emitters at $z \approx 3.3$ in the field.
The Keck Cosmic Web Imager (KCWI) \citep{Morrissey2018} integral field spectrograph with its bluer wavelength coverage has also been used for similar studies like the CGM at Cosmic Noon with KCWI program \citep{Nielsen2020}.

Amongst the possible absorption lines observed towards bright background QSOs, the Ly-$\alpha$ 1215 \AA\ line is an important tracer of neutral hydrogen (\ion{H}{i}), the fundamental building block of stars in galaxies. 
Several surveys have targeted Ly-$\alpha$ absorbers at an intermediate redshift range $0.4 \lesssim z \lesssim 2$ such as the Cosmic Ultraviolet Baryons Survey \citep[CUBS,][]{Chen2020} and the Bimodal Absorption System Imaging Campaign \citep[BASIC,][]{Berg2022}. 
The CUBS survey targeted 15 QSOs selected by their near ultraviolet (NUV) brightness to search primarily for systems with column densities \logNHIunit $< 19.0$. 
Galaxies associated with these low column density systems vary significantly in properties such as SFR, metallicity and environment \citep{Chen2020, Cooper2021}. 
Similarly, the BASIC survey targeted a sample of 36 Lyman limit systems (LLS) and pseudo-LLS absorbers at $z\lesssim 1$ and found no associated galaxies for the majority of absorbers with $< 10$ per cent solar metallicity, pointing to a population of low-metallicity absorbers not associated with the CGM \citep{Berg2022}.

Tangential to these surveys is the MUSE-ALMA Halos Survey where 32 \ion{H}{i} Ly-$\alpha$ absorbers with column densities ranging from $18.1 \lesssim \logNHI \lesssim 21.7$ at redshift $0.2 \lesssim z \lesssim 1.4$ are targeted \citep[see][for an overview]{Peroux2022}. 
The targets were selected from known quasar absorbers with measured \ion{H}{i} column densities $\logNHIunit > 18.0$ from HST UV spectroscopy. 
Early findings from the survey include the discovery of gas likely tied to the intragroup medium \citep{Peroux2017, Peroux2019}, accretion \citep{Rahmani2018a} and galactic winds \citep{Rahmani2018b}. 
Both \citet{Klitsch2018} and \citet{Szakacs2021} combine MUSE data with Atacama Large Millimetre/submillimetre Array (ALMA) data to investigate the molecular and ionized gas properties of galaxies at the same redshift as the \ion{H}{i} absorbers. 
The molecular gas in associated galaxies extends further than the ionized disk \citep{Klitsch2018} and is kinematically coupled with the ionized gas \citep{Szakacs2021}.
\citet{Hamanowicz2020} investigated correlations between absorber and galaxy properties for a sub-sample of 14 absorbers.
Our work is a continuation of that study for the full MUSE-ALMA Halos survey. 
With this complete sample of 32 absorbers, we aim to examine the distribution of gas and metals in the CGM, and how this distribution is related to the properties of associated galaxies. 
We adopt the following cosmology: $\rm H_0 = 70$ \kms $\rm {Mpc}^{-1}$, $\Omega_M = 0.3$ and $\Omega_\Lambda = 0.7$.

\section{New MUSE Observations}
\label{sec: obs}
The VLT/MUSE observations presented in this paper were carried out across various programmes 
(ESO 96.A-0303, 100-A-0753, 101.A-0660 and 102.A-0370, PI: C. Péroux and 298.A-0517, PI: A. Klitsch).
For further details on the observing strategy, data reduction and data quality checks (astrometry, wavelength and flux calibration), we refer the reader to \citet{Peroux2022}. 
For the sake of completeness, we summarise the detection methods here:
\begin{enumerate}
    \item 
    Passive galaxies without detectable emission were found with a continuum search using the \textsc{ProFound} \footnote{https://github.com/asgr/ProFound} tool \citep{Robotham2018}. 
    The limiting continuum magnitudes are calculated within an aperture with radius equal to the seeing in a pseudo $r$-band image of the MUSE cube.
    Our calculated 3$\sigma$ limits in \autoref{tab:abs} vary between MUSE fields because of the different exposure times and seeing conditions. 
    \item
    In addition to continuum sources, we expect a population of faint objects only detectable by their emission lines. 
    To systematically search for these objects without continuum, we used the MUSE Line Emission Tracker (MUSELET) module of the MUSE Python Data Analysis Framework (\textsc{MPDAF}) package \footnote{https://mpdaf.readthedocs.io/en/latest/index.html\#} \citep{MPDAF2016}. 
    For each absorber, we calculate the $3\sigma$ star-formation rate limit. 
    We assume the source is unresolved over a disk with diameter equal to the seeing and an emission line with full width at half maximum (FWHM) of 3 \AA.
    From this SFR limit, we estimate a stellar mass limit assuming the galaxy is on the main sequence \citep{Schreiber2015}.
    These values are tabulated in \autoref{tab:abs}.
    \item
    We also use complementary Hubble Space Telescope (HST) imaging from a 40-orbit medium programme (ID: 15939, PI: Peroux) and additional archival HST data. 
    The fainter detection limit in the HST broadband images enabled us to search for objects not detected by previous methods. 
    This approach contributed 10 additional sources which were typically [\ion{O}{ii}] or Ly-$\alpha$ emitters missed by MUSELET. 
    All continuum galaxies identified in the MUSE data were recovered in the HST data if the field was observed with a red HST filter (e.g. F702W and F814W).
    \item
    Finally, we perform a spectral PSF subtraction of the QSO to search for objects at low separation in the MUSE cube using \textsc{QFitsView}\footnote{https://www.mpe.mpg.de/~ott/QFitsView/} \citep{QFitsView}. 
    This search was performed at the redshifts of \ion{H}{i} absorbers to ensure no emission-line galaxies near the QSO were missed. 
    The linemaps at the expected position of strong emission lines are created surrounding the background QSO which effectively subtracts out the QSO continuum. 
    A galaxy obscured by the QSO point spread function (PSF) in the field Q0454+039 was detected in this manner.
\end{enumerate}

\begin{figure}
    \includegraphics[width=\columnwidth]{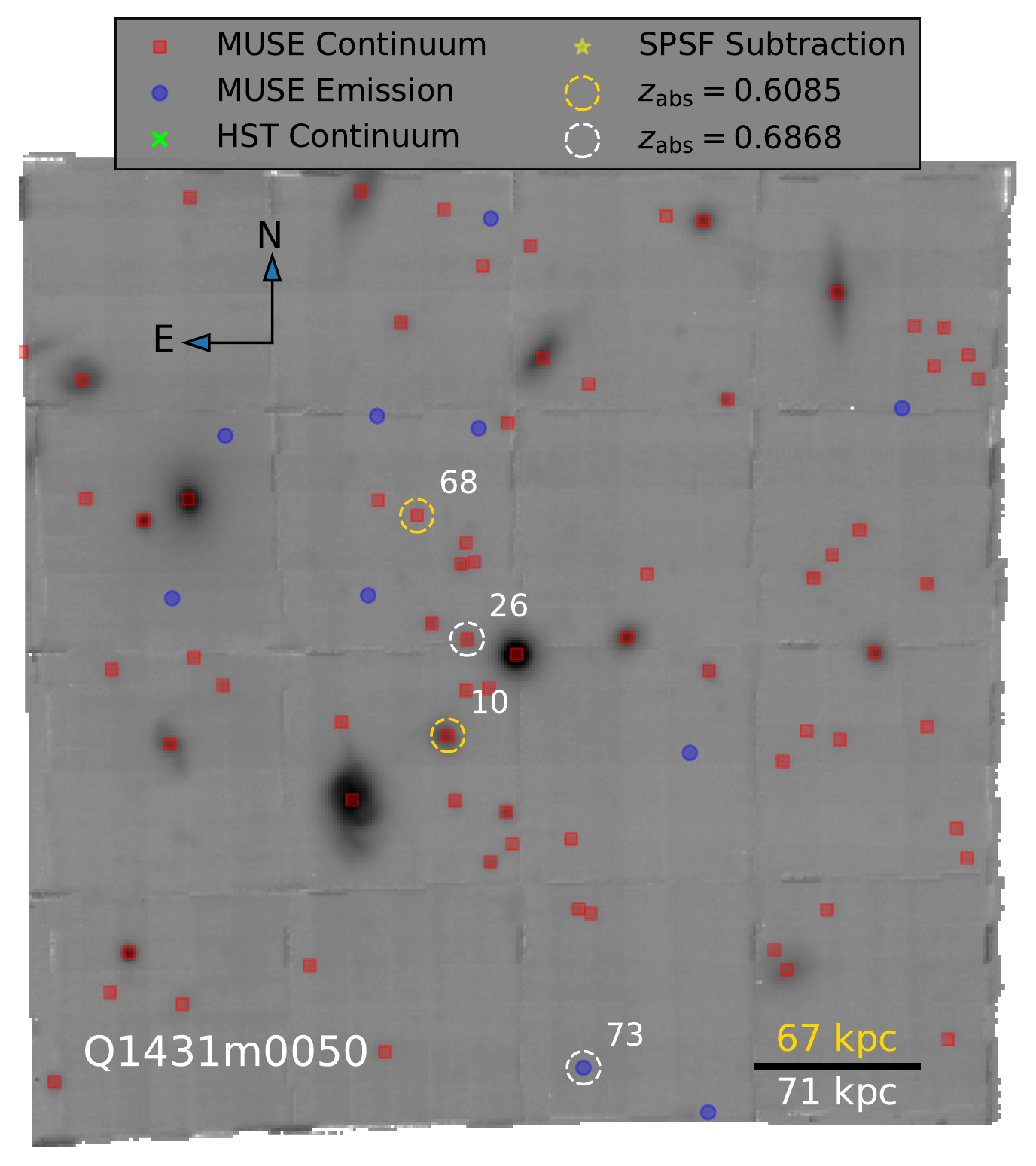}
    \caption{
    A white-light image of the MUSE field centred on quasar Q1431$-$0050. 
    The four markers represent how the object was detected. 
    Red squares denote continuum objects detected using \textsc{ProFound} and blue circles are emission-line sources. 
    Green crosses are objects found after we used sources in the HST continuum imaging as a prior that were not found initially during our MUSE continuum and emission-line searches. 
    Gold stars represent objects detected using a spectral PSF subtraction of the quasar. 
    No objects are detected in the Q1431$-$0050 field using HST images as a prior and after a QSO subtraction, but we direct the reader to Figures \ref{fig:Fields1}, \ref{fig:Fields2} and \ref{fig:Fields3} for such instances.
    We note that many sources will be detected in multiple ways; an emission-line galaxy with bright stellar continuum will be found in MUSE continuum, emission and in the HST continuum imaging. 
    The markers only denote how the source was first detected, following the order of methods listed in \autoref{sec: obs} (using a MUSE continuum and emission-line search, HST continuum detections as a prior and a spectral PSF subtraction of the QSO).
    The dashed circles are galaxies that are within $\pm 500$ \kms of an absorber and they are labelled by their ID in the MUSE-ALMA Halos object catalogues \citep{Peroux2022}.
    Ten arcseconds corresponds to 67 and 71 kpc at $z_{\rm abs} = 0.6085$ and $z_{\rm abs} = 0.6868$ respectively.
    North is up and east is left.}
    \label{fig:Q1431}
\end{figure}

\autoref{fig:Q1431} shows a white-light image of the Q1431$-$0050 field with all objects marked by how they were first detected. 
The dashed circles represent galaxies associated with the absorber at the redshift found in the legend. 
Similar white-light images of the remaining fields can be found in Appendix \ref{App:whitelight}.

Redshifts for all objects detected in the MUSE data were determined using the \textsc{marz} \footnote{https://samreay.github.io/Marz/\#/overview} tool \citep{Hinton2016} and independently verified by two experts (SW and CP).
We refer the reader to \citet{Peroux2022} for a more detailed methodology.

\begin{table*}
\begin{center}
\caption{{\bf Summary of absorber properties targeted in the MUSE-ALMA Halos Survey.}
Absorber redshifts, \ion{H}{i} column densities and metal abundances are tabulated. 
We also include the survey area corresponding to the 60 arcsec $\times$ 60 arcsec field-of-view (FoV) of MUSE at the absorber redshift.
For each field and absorber redshift, we provide $3\sigma$ star-formation rate and equivalent stellar mass limits using the SFR-$M_*$ main sequence. 
The number of galaxies within $\pm 500$ \kms of the quasar absorbers is also given in the penultimate column.
The final column contains references for the absorber \ion{H}{i} column density and metal abundance.
}
\begin{tabular}{llcccccccccl}
\hline\hline
Quasar &$z_{\rm abs}$ &$\logNHI$ & [Zn/H] & [Fe/H] & Survey Area & $m_r$ limit & SFR limit & $M_*$ limit & $N_{\rm gal}$ & Reference \\
       &            & (cm$^{-2}$) & $12 + \log$(O/H) &  $12 + \log$(O/H) & (kpc $\times$ kpc) & mag & (M$_{\odot}$ yr$^{-1}$) & ($\log \ $M$_{\odot}$) & &  \\
\hline                                                        
Q0058$+$0019 &0.6125 &20.08$\pm$0.15 & $0.08 \pm 0.21$ & $-0.01 \pm 0.16$ & 405 $\times$ 405 & 24.0 & $7.3 \times 10^{-1}$ & 9.1 & 0 & P00, P00 \\
Q0123$-$0058 &0.8686 &$<$18.62 & -- & -- & 463 $\times$ 463 & 23.2 & $3.1 \times 10^{0}$ & 9.6 & 0 & R06 \\
...          &1.4094 &{20.08$\pm$0.09} & $-0.45 \pm 0.2$ & $-0.25 \pm 0.12$ & 506 $\times$ 506 & ... & $1.4\times 10^{1}$ & 10 & 0 & M09, M09\\
Q0138$-$0005 &0.7821 &{19.81$\pm$0.08} & $0.28 \pm 0.16$ & $> 0.09$ & 447 $\times$ 447 & 23.5 & $1.6 \times 10^{1}$ & 9.3 & 1 & P08, P08 \\
J0152$-$2001 &0.3830&{$<$18.78} & -- & $> -1.36$ & 314 $\times$ 314 & 25.6 & $2.8 \times 10^{-2}$ & 7.7 & 6 & R18, H20 \\
...          &0.7802 &{18.87$\pm$0.12} & -- & $> -0.44$ & 447 $\times$ 447 & ... & $1.9 \times 10^{-1}$ & 8.4 & 1 & R06, H20 \\
Q0152$+$0023 &0.4818 &{19.78$\pm$0.08} & -- & -- & 359 $\times$ 359 & 24.6 & $2.4 \times 10^{-1}$ & 8.6 & 4 & R06 \\
Q0420$-$0127 &0.6331 &{18.54$\pm$0.09} & -- & -- & 411 $\times$ 411 & 25.2 & $1.1 \times 10^{-1}$ & 8.2 & 4 & R06 \\ 
Q0454$+$039  &0.8596 &{20.69$\pm$0.03} & $-1.01 \pm 0.12$ & $-0.73 \pm 0.08$ & 356 $\times$ 356 & 26.5 & $1.2 \times 10^{-1}$ & 8.3 & 2 & P00, P00 \\
...          &1.1532 &{18.59$\pm$0.02} & -- & -- & 360 $\times$ 360 & ... & $2.7 \times 10^{-1}$ & 8.4 & 2 & R06 \\
Q0454$-$220  &0.4744 &{19.45$\pm$0.03} & -- & -- & 461 $\times$ 461 & 25.3 & $1.0 \times 10^{-1}$ & 8.3 & 1 & R06 \\
...          &0.4833 &{18.65$\pm$0.02} & -- & -- & 495 $\times$ 495 & ... & $1.1 \times 10^{-1}$ & 8.3 & 1 & R06 \\
Q1110$+$0048 &0.5604 &{20.20$\pm$0.10} & -- & -- & 388 $\times$ 388 & 26.2 & $1.2 \times 10^{-1}$ & 8.3 & 3 & R06 \\
J1130$-$1449 &0.1906 &{$<$19.10}       & -- & -- & 191 $\times$ 191 & 25.7 & $4.0  \times 10^{-3}$ & 7.0 & 1 & R06 \\
...          &0.3127 &{21.71$\pm$0.08} & -- & $-1.64 \pm 0.08$ & 275 $\times$ 275 & ... & $1.8 \times 10^{-2}$ & 7.6 & 13 & R06, P19 \\
...          &0.3283 &{$<$18.90}       & -- & -- & 284 $\times$ 284 & ... & $1.2 \times 10^{-2}$ & 7.4 & 2 & R06 \\
J1211$+$1030 &0.3929 &{19.46$\pm$0.08} & -- & $> -1.05$ & 319 $\times$ 319 & 25.3 & $3.5 \times 10^{-2}$ & 7.8 & 3 & R06, H20 \\
...          &0.6296 &{20.30$\pm$0.24} & -- & $-0.68 \pm 0.3$ & 410 $\times$ 410 & ... & $2.2 \times 10^{-1}$ & 8.5 & 2 & R06, H20 \\
...          &0.8999 &{$<$18.50}       & -- & -- & 467 $\times$ 467 & ... & $3.9 \times 10^{-1}$ & 8.7 & 5 & R06 \\
...          &1.0496 &{$<$18.90}       & -- & $> 1.69$ & 486 $\times$ 486 & ... & $7.6 \times 10^{-1}$ & 8.9 & 0 & R06, H20 \\
Q1229$-$021  &0.3950 &{20.75$\pm$0.07} & $-0.5$ & $< -1.31$ & 320 $\times$ 320 & 24.9 & $4.4 \times 10^{-2}$& 7.9 & 1 & B98, B98\\
...          &0.7572 &{18.36$\pm$0.09} & -- & $> -1.48$ & 442 $\times$ 442 & ... & $3.4 \times 10^{-1}$ & 8.7 & 1 & R06, H20 \\
...          &0.7691 &{18.11$\pm$0.15} & -- & $> -2.34$ & 444 $\times$ 444 & ... & $3.4 \times 10^{-1}$ & 8.7 & 3 & R06, H20 \\
...          &0.8311 &{18.84$\pm$0.10} & -- & $> -2.19$ & 456 $\times$ 456 & ... & $3.7 \times 10^{-1}$ & 8.7 & 4 & R06, H20 \\
Q1342$-$0035 &0.5380 &{19.78$\pm$0.13} & -- & -- & 380 $\times$ 380 & 24.1 & $4.4 \times 10^{-1}$& 8.9 & 2 & R06 \\
Q1345$-$0023 &0.6057 &{18.85$\pm$0.20} & -- & -- & 403 $\times$ 403 & 25.4 & $1.9 \times 10^{-1}$ & 8.5 & 2 & R06 \\
Q1431$-$0050 &0.6085 &{19.18$\pm$0.24} & -- & -- & 404 $\times$ 404 & 25.1 & $2.6 \times 10^{-1}$ & 8.6 & 3 & R06 \\
...          &0.6868 &{18.40$\pm$0.07} & -- & -- & 425 $\times$ 425 & ... & $3.7 \times 10^{-1}$ & 8.7 & 2 & R06 \\
Q1515$+$0410 &0.5592 &{20.20$\pm$0.19} & $< 0.64$ & -- & 388 $\times$ 388 & 25.5 & $1.7 \times 10^{-1}$ & 8.4 & 4 & R16, R16 \\
Q1554$-$203  &0.7869 &{$<$19.00}       & -- & -- & 448 $\times$ 448 & 25.3 & $2.4 \times 10^{-1}$& 8.5 & 1 & R06 \\
J2131$-$1207 &0.4298 &{19.50$\pm$0.15} & -- & $> -0.96$ & 337 $\times$ 337 & 25.6 & $7.1 \times 10^{-2}$ & 8.1 & 4 & M16, H20 \\
Q2353$-$0028 &0.6044 &{21.54$\pm$0.15} & $-0.92 \pm 0.32$ & -- & 402 $\times$ 402 & 25.6 & $1.8 \times 10^{-1}$ & 8.5 & 0 & R06, N08 \\
\hline\hline 				       			 	 
\label{tab:abs}
\end{tabular}			       			 	 
\begin{minipage}{180mm}
  {B98: \citet{Boisse1998},
  H20: \citet{Hamanowicz2020},
  M09: \citet{Meiring2009},
  M16: \citet{Muzahid2016}
  N08: \citet{Nestor2008},
  P00: \citet{Pettini2000},
  P08: \citet{Peroux2008},
  P19: \citet{Peroux2019},
  R06: \citet{Rao2006},
  R16: \citet{Rahmani2016},
  R18: \citet{Rahmani2018b},
  }
\end{minipage}
\end{center}			       			 	 
\end{table*}	

\section{Associated Galaxy Properties}
\subsection{Identification of galaxies associated with absorbers}
The selection criteria for galaxies that are `associated' with the \ion{H}{i} Ly-$\alpha$ absorbers are critical to our study. 
We require galaxies to be within $\pm 500$ \kms of the absorber. 
This selection differs from previous survey papers \citep[e.g.][uses a $\pm 1000$ \kms window and includes three additional galaxies]{Hamanowicz2020} because recent studies have shown that \ion{Mg}{ii} absorbers that trace dense, neutral gas are rarely detected beyond $\pm 300$ \kms \citep{Huang2021}. 
EAGLE simulations of more spatially extended ionized gas (\ion{O}{vi} $\lambdaup \lambdaup 1032, 1038$) find that line-of-sight velocity cuts of $\pm 500$ \kms include significant amounts of gas beyond the virial radius \citep{Ho2021}. 
We expect the host galaxies of the dense and neutral absorbers in the MUSE-ALMA Halos survey to be within a smaller velocity window than the \ion{O}{vi} but maintain a velocity cut of $\pm 500$ \kms to be consistent with other surveys \citep[][]{Dutta2020}. 
However, we note that other studies have used different criteria \citep[e.g.][]{Schroetter2016, Berg2022}.
Absorber redshifts are tabulated in \autoref{tab:abs} and galaxy redshifts are obtained from the final catalogues of the MUSE-ALMA Halos Survey. 
The typical uncertainty in the galaxy redshift is $50$--$60$ \kms\!.
The velocity difference between the galaxy and the absorber is calculated using $\Delta v =  c(z_{\rm gal} - z_{\rm abs})/(1 + z_{\rm abs})$. 
We find 79 galaxies within $\pm 500$ \kms of 27 absorbers and we henceforth refer to this sample as associated galaxies. 
Included in this sample are six associated galaxy candidates that have low signal-to-noise ratio (SNR) emission or absorption features near the absorber redshifts.
Five absorbers do not have galaxy counterparts.
We note that 75 per cent of objects are found within $\sim$200 \kms of the absorber and $90$ per cent within $\sim$350 \kms\!.
If we restrict associated galaxies to be within $\pm 300$ \kms of absorbers, we still find that 27/32 absorbers have an accompanying galaxy.

In \autoref{tab:abs}, we list the properties of the absorber and the number of associated galaxies ($N_{\rm gal}$) found for each absorber.
We flag the redshifts of the possible galaxy counterparts with quality 2 in \autoref{tab:abs} to be consistent with the catalogues in \citet{Peroux2022}.
For five absorbers in four fields, we do not detect any galaxies within $\pm 500$ \kms\!.

\subsection{Spectral Extraction}
The circular apertures used to extract spectra were checked to determine whether all the emission flux was captured.
For sources with extended emission, the apertures were re-drawn using the segmentation maps produced by \textsc{ProFound} or manually on \textsc{QFitsView} using linemaps. 
This additional step was only necessary for three sources and ensures we have accurate measurements of the emission-line flux to determine properties such as SFR and metallicity.

\subsection{Flux Measurements}
Initially, emission lines [\ion{O}{ii}] $\lambdaup \lambdaup 3726, 3729$, H$\delta$ $\lambdaup 4102$, H$\gamma$ $\lambdaup 4340$, H$\beta$ $\lambdaup 4861$, [\ion{O}{iii}] $\lambdaup \lambdaup 4959, 5007$, [\ion{N}{ii}] $\lambdaup \lambdaup 6549, 6583$, H$\alpha$ $\lambdaup 6563$ and [\ion{S}{ii}] $\lambdaup \lambdaup 6717, 6730$ were automatically fitted using the \textsc{MPDAF} 1D Gaussian function \citep{Bacon2016}. 
We inspected the quality of the fits and searched for cases where the continuum estimation (using a width of $40$ pixels centred on the line) appeared incorrect. 
This was typically caused by emission lines being near the GALACSI Adaptive Optics (AO) gap centred on the rest-frame \ion{Na}{i} D line, or the presence of strong sky residuals. 
Lines near the AO gap required a custom and asymmetric window for continuum estimation while sky residuals were masked during the refit.

\begin{figure*}
    \includegraphics[width=\textwidth]{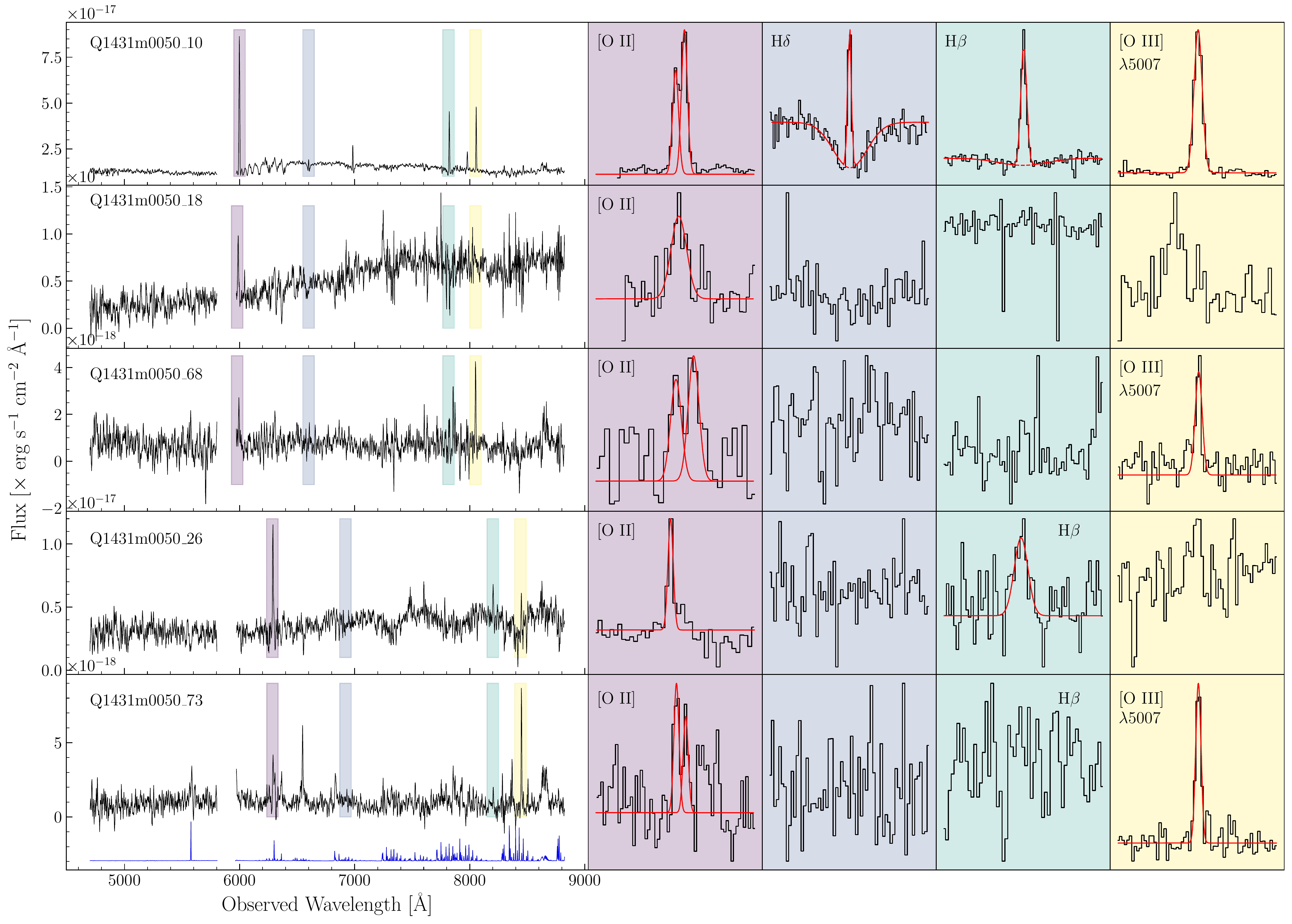}
    \caption{The spectra of galaxies associated with absorbers in the Q1431$-$0050 field and corresponding emission-line fits. 
    The left panels show the MUSE spectra of the associated galaxies after smoothing with a 5-pixel boxcar and the galaxy identification from the MUSE-ALMA Halos catalogues is displayed in the top left.
    In the bottom-left panel, a sky spectrum is included in blue.
    The right panels show our [\ion{O}{ii}], H$\delta$, H$\beta$ and [\ion{O}{iii}] $\lambdaup 5007$ emission-line fits in red to unsmoothed spectra.
    The background colours correspond to the boxed emission lines in the full spectra.
    We present a variety of different emission-line fits here: a double Gaussian for [\ion{O}{ii}] (e.g. objects Q1431m0050\_10 and  Q1431m0050\_68) and emission with stellar absorption (two centre panels for object Q1431m0050\_10).
    We also display non-detections of emission lines in the spectra.
    }
    \label{fig:Q1431spec}
\end{figure*}

Broad stellar absorption around the Balmer lines was fitted out using the method from \citet{Zych2007}, which involved simultaneously fitting the absorption and emission using two Gaussians.
For the resolved [\ion{O}{ii}] doublet, a double Gaussian fit was used. 
Examples of these fits for associated galaxies in the Q1431$-$0050 field are shown in \autoref{fig:Q1431spec}. 

For galaxies at redshift $z \lesssim 0.42$, the observed Balmer decrement H$\alpha/$H$\beta$ can be used to determine the dust attenuation. 
The intrinsic flux ratio of these lines is expected to be H$\alpha/$H$\beta = 2.87$ \citep{Osterbrock1989} assuming case-B recombination and any deviation from this value is expected to be caused by dust. 
The colour excess $E(B-V)$ was calculated assuming a Small Magellanic Cloud-type (SMC-type) extinction curve \citep{Pei1992} using the prescription in \citet{Gharanfoli2007}. 
It has been found that the average extinction curve in \ion{Mg}{ii} systems at redshift $1 \leq z_{\rm abs} < 2$ resembles a SMC-type curve \citep{York2006}.
Uncertainties were determined by propagating errors in the H$\beta$ and H$\alpha$ flux measurements.  

\section{Relating Absorber and Galaxy Properties}
\subsection{Distribution of Galaxies}
With the associated galaxy sample finalised, we can begin investigating the relationships between CGM gas properties such as \ion{H}{i} column density and the properties of the associated galaxies. 
Early studies connecting \ion{H}{i} absorbers with nearby galaxies typically found a single associated object \citep[e.g.][]{Bergeron1986a, BergeronBoisse1991}. 
Later works that build significant samples of absorber-galaxy systems use long-slit or multi-slit spectroscopy on objects seen in imaging of the QSO field \citep[e.g.][]{Chen2009, Thom2011, Ribaudo2011, Lehner2013}. 
These methods bias against detections of galaxies at very small angular separations to the QSO. 
In 7 out of the 19 fields (e.g. Q0454+039 at $z_{\rm abs} = 0.8596$ and Q1110+0048 at $z_{\rm abs} = 0.5604$), we find galaxies within three arcseconds of the QSO. 
Many of these earlier works that rely on imaging and then spectroscopy are incomplete in this regard.
Moreover, galaxies with faint continuum will also be missed in imaging; the ability of integral field spectroscopy to simultaneously search for continuum and emission objects has greatly improved the completeness of searches for absorber-galaxy pairs and recent surveys often find that multiple galaxies can be associated with a single absorber \citep{Hamanowicz2020, Chen2020, Berg2022}. 

We also note there are various targeted surveys of the CGM surrounding isolated galaxies \citep[e.g.][]{Tumlinson2013, Liang2014, Borthakur2015, Prochaska2017, Berg2018, Chen2018, Pointon2019, Kulkarni2022}. 
These `galaxy-centric' surveys cross-correlate background quasar sightlines with foreground galaxies pre-selected by their properties. 
In this work, we choose to plot results from `absorber-centric' surveys, where the pre-selection is based on the absorber properties and the types of galaxies and environments studied vary more significantly.

\begin{table*}
\begin{center}
\caption{\textbf{Summary of associated galaxy properties.}
The identifications of galaxies within $\pm 500$ \kms of an absorber. 
Velocity differences ($\Delta v$) are calculated with respect to the absorber redshift. 
Star-formation rates and limits are based on \citet{Kennicutt1998}, and we use the H$\alpha$ empirical relation when available. 
At $z \gtrsim 0.4$, we use the [\ion{O}{ii}] $\lambdaup \lambdaup 3726, 3729$ luminosity to estimate the SFR. 
Dust corrections are performed for galaxies with measured H$\beta$ and H$\alpha$ emission-line fluxes.
The tabulated values assume a Salpeter initial mass function \citep[IMF;][]{Salpeter1955} and choosing to adopt the Chabrier IMF \citep{Chabrier2003} will reduce the SFRs by a factor of 1.8. 
Both $\rm R_3$ lower and upper branch metallicites derived from \citet{Curti2017, Curti2020a} are provided when available. 
The extinction calculated using the Balmer decrement and galaxy type (star-forming or passive) are also provided.
Galaxies are classed as star-forming (SF) if emission lines are seen in the spectrum and passive (P) if not. 
The redshift confidence (RC) is also given where a value of $2$ indicates a possible associated galaxy due to low SNR absorption or emission lines. 
A value of $3$ is a likely redshift where multiple high SNR lines are detected.
Entries with value $-999$ correspond to measurements that are unavailable. 
For example, corrected SFRs are unavailable for galaxies without detectable or measurable H$\beta$ and H$\alpha$ emission.
A complete table of associated galaxies and their properties is available online.}
\begin{tabular}{llccccccccc}
\hline\hline
ID &$z_{\rm gal}$ & $b$ & $\Delta v$ & SFR & SFR$_{\rm corr}$ & 12 + $\log$(O/H)$_{\rm lower}$ & 12 + $\log$(O/H)$_{\rm upper}$ & $E(B-V)$ & Type & RC\\
    &         &  (kpc) & ($\kms$\!) & (M$_\odot$ yr$^{-1}$) & (M$_{\odot}$ yr$^{-1}$) & & & & &  \\
\hline
\multicolumn{11}{c}{\bolden{Q$0138-0005$, $z_{\rm quasar} = 1.96$, $z_{\rm abs} = 0.7821$, $\logNHIunit = 20.08 \pm 0.15$}} \\
\hline
Q0138m0005\_14 & 0.7821 & 82 & -3.4 & $6.9 \pm 2.4$ & -999 & -999 & $8.7 \pm 0.1 $ & -999 & SF & 3 \\
\hline
\multicolumn{11}{c}{\bolden{Q$0152-2001$, $z_{\rm quasar} = 2.06$, $z_{\rm abs} = 0.3830$, $\logNHIunit < 18.78$}} \\
\hline
Q0152m2001\_4  & 0.3814 & 180 & 340 & $< 0.06$ & -999 & -999 & $8.6 \pm 0.2 $ & -999 & P & 3\\
Q0152m2001\_5  & 0.3826 & 60 & 84 & $0.73 \pm 0.4$ & $2.2 \pm 0.5$ & -999 & $8.6 \pm 0.1$ & $0.53 \pm 0.04$ & SF & 3 \\
Q0152m2001\_7  & 0.3814 & 150 & 350 & $0.18 \pm 0.4$ & -999 & -999 & $<8.7$ & -999 & SF & 3\\
Q0152m2001\_13 & 0.3815 & 84 & 330 & $0.10 \pm 0.4$ & $0.1 \pm 0.4$ & -999 & $8.7 \pm 0.1$ & $0.019 \pm 0.002$ & SF &3 \\
... & ... & ... & ... & ... & ... & ... & ... & ... & ... & ...  \\
\hline\hline 				       			 	 
\label{tab:associated}
\end{tabular}			       			 	 
\begin{minipage}{180mm}

\end{minipage}
\end{center}			       			 	 
\end{table*}	

\begin{figure}
    \includegraphics[width=\columnwidth]{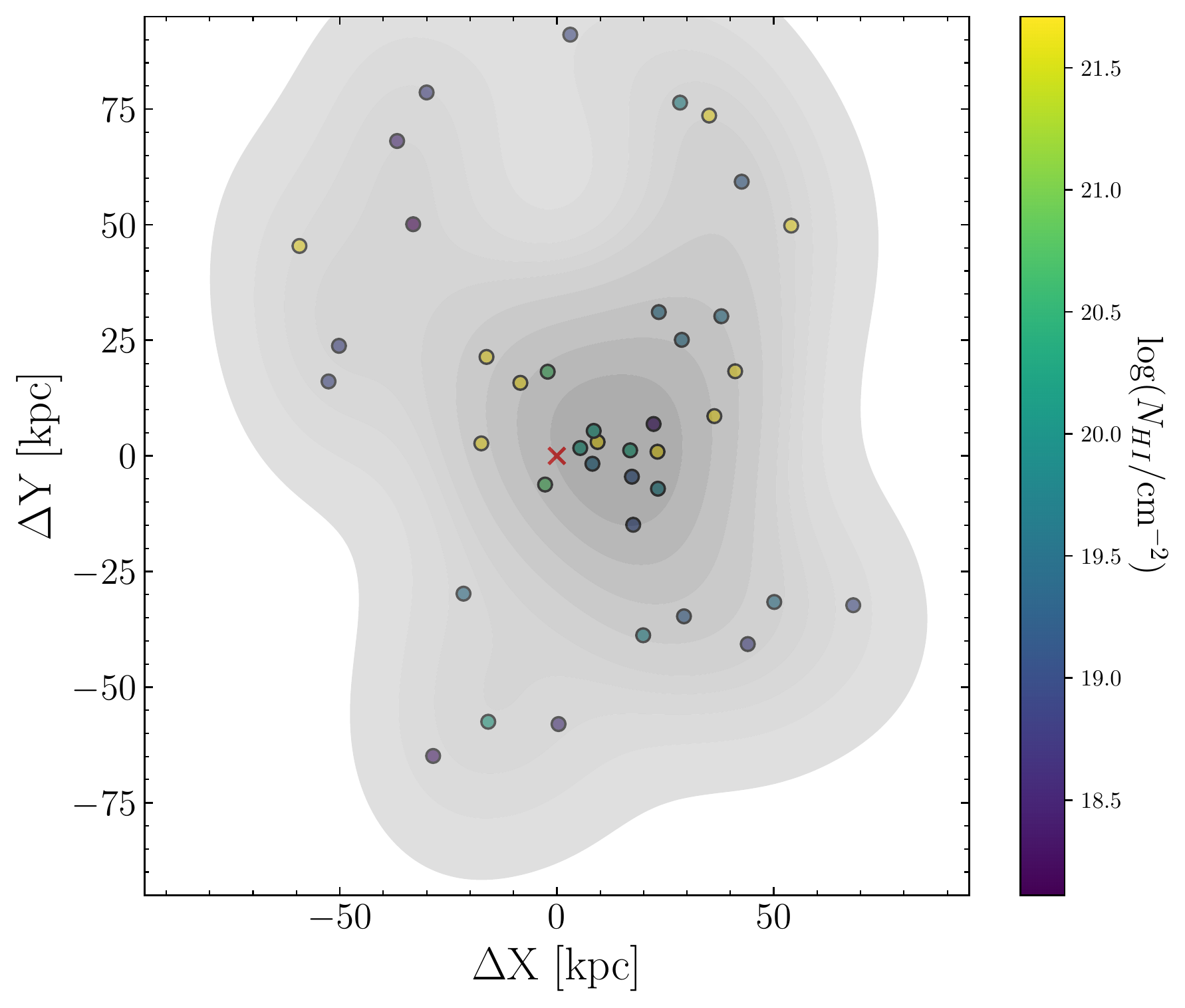}
    \caption{The distribution of associated galaxies around absorbers using their physical distances from the absorber (red cross). 
    Each galaxy is coloured by the \ion{H}{i} column density of the absorber they are associated with. 
    We use a two-dimensional Gaussian kernel density estimator (KDE) to construct a surface density map of the associated galaxies. 
    The darker areas represent regions of high density. 
    We limit the associated galaxies plotted to those within the area covered by the MUSE field-of-view of the lowest redshift absorber ($z_{\rm abs} = 0.1906$ corresponding to 191 kpc $\times$ 191 kpc). 
    There are 41 galaxies within this area.
    }
    \label{fig:RA_Dec_dist}
\end{figure}

In \autoref{fig:RA_Dec_dist}, we plot the distribution of associated galaxies around absorbers in physical space. 
We only include galaxies found within a 191 kpc $\times$ 191 kpc region centred on the QSO. 
This corresponds to the MUSE areal coverage of the $z = 0.1906$ absorber and reduces the sample of associated galaxies to 41.
The overlaid two-dimensional kernel density estimation reveals the probability of finding an associated galaxy is higher at close separations to the absorber. 
The surface density map is not centred on the absorber (red cross), but this is likely caused by the variability in small samples. 

\begin{figure}
    \includegraphics[width=\columnwidth]{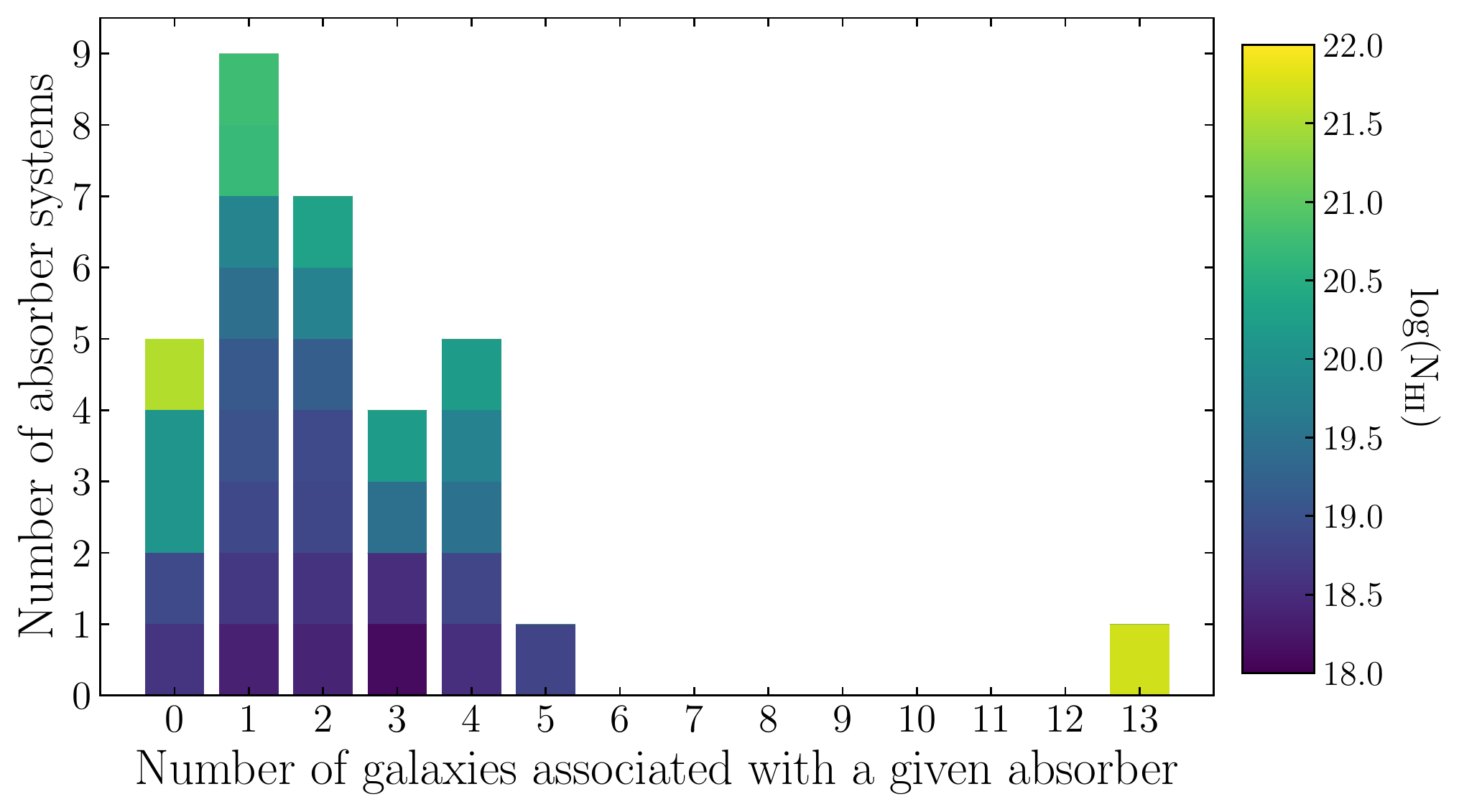}
    \caption{The number of associated galaxies ($N_{\rm gal}$) identified per absorber. 
    Each absorber is colour-coded by its \ion{H}{i} column density. 
    17 out of the 32 absorbers (53 per cent) with MUSE IFS observations have at least two associated galaxies.
    It appears that there is no trend between $N_\ion{H}{i}$ and the number of associated galaxies.
    }
    \label{fig:N_N_gal_NHI}
\end{figure}

\autoref{fig:N_N_gal_NHI} reveals that 17 out of the 32 (59 per cent) absorbers have two or more objects within $\pm 500$ \kms of the neutral gas. 
We also find that the distribution of column densities varies significantly for each bin of the number of associated galaxies. 
While we expect high \ion{H}{i} column density systems to be associated with galaxies at smaller impact parameters \citep{Peroux2005, Noterdaeme2014}, it remains unclear whether the environment of absorbers depends on $N(\ion{H}{i})$. 
Recent works using multiple QSO sightlines or an extended background source reveal $> 1$ dex changes in column density across separations of only several kpc \citep{Kulkarni2019, Augustin2018, Cashman2021, Bordoloi2022}, suggesting $N(\ion{H}{i})$ likely does not correlate with the number of associated galaxies because gas in the CGM is inhomogeneously distributed.  

\begin{figure}
    \includegraphics[width=\columnwidth]{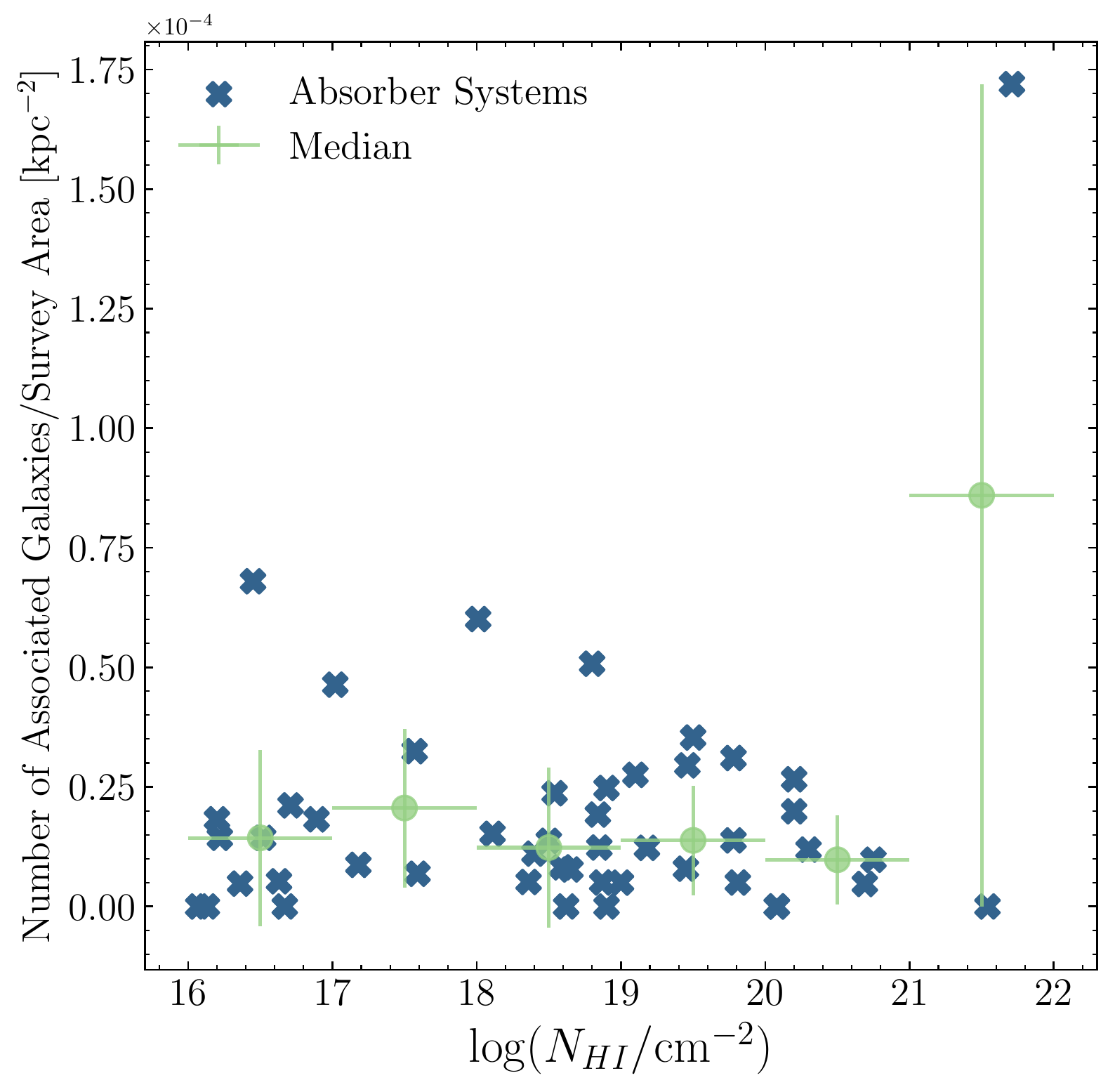}
    \caption{The number of galaxies associated with a given absorber ($N_{\rm gal}$) normalised by the survey area plotted against the absorber column density. 
    We include in this plot absorbers analysed with MUSE data from the CUBS and BASIC surveys.
    The green points give the median value in 1 dex bins of \ion{H}{i} column density from $\logNHIunit = 16.0 - 22.0$. 
    The error in the median value of $N_{\rm gal}/$area is calculated using the standard deviation of values within a given bin.
    We find that this value does not evolve with the column density of the absorber. 
    The large median value in the highest column-density bin is attributed to a 13-member galaxy group associated with a single absorber at $z_{\rm abs} = 0.313$.
    }
    \label{fig:N_gal_NHI}
\end{figure}

To test this, in \autoref{fig:N_gal_NHI} we plot the number of associated galaxies normalised by the physical area of the MUSE FoV found for a given absorber column density. 
We expand the MUSE-ALMA Halos sample to lower column densities using the CUBS \citep{Chen2020} and BASIC \citep{Berg2022} surveys and only include absorbers observed with MUSE. 
The median $N_{\rm gal}$ value using 1 dex bins in the range $16.0 < \logNHIunit < 22.0$ is plotted in red.
There appears to be no correlation between the column density of the \ion{H}{i} and the number of associated galaxies detected after controlling for the different physical scales probed by MUSE at the various absorber redshifts.
While we find a rich galaxy group associated with our highest column density system ($\logNHIunit = 21.71 \pm 0.08$ for $z_{\rm abs} = 0.3127$ in J1130-1449), there is no galaxy near the Q2353-0028 absorber with \logNHIunit$ = 21.54 \pm 0.15$ and $z_{\rm abs} = 0.6044$ down to a SDSS $r$-band magnitude of $\sim$25 and SFR limit of $0.18$ \Moyr.
The large median value in the final bin is caused by the significant overdensity in the former system and we require more observations of \logNHIunit $> 21.0$ absorbers for a statistically significant result.

\begin{figure}
    \includegraphics[width=\columnwidth]{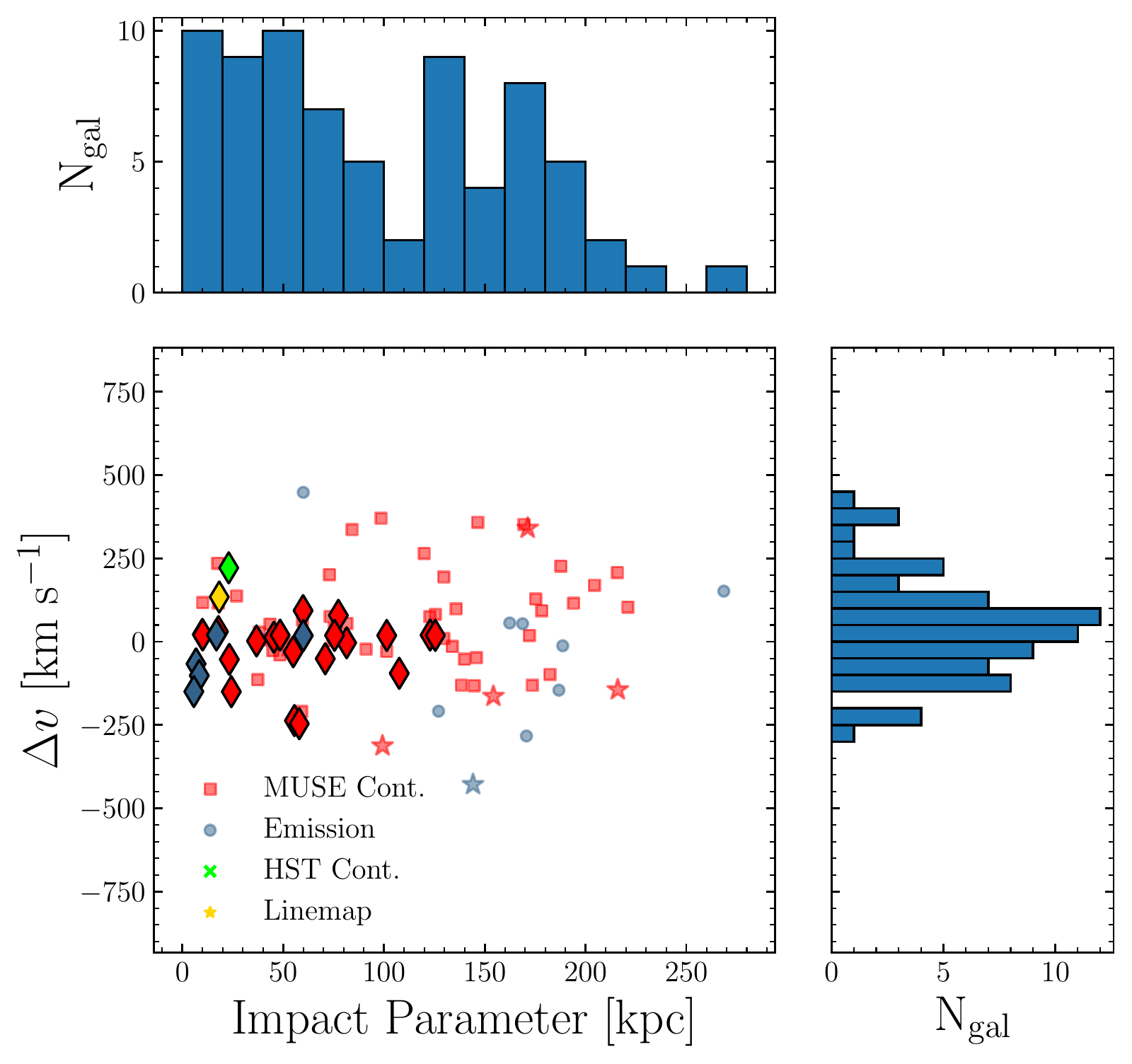}
    \caption{The distribution of associated galaxies in impact parameter and velocity space. 
    The galaxy velocities appear to peak near the absorber redshift, with an extended tail at large positive velocities. 
    More galaxies are found at low impact parameters, but this effect may be caused by the MUSE FoV limiting the areal coverage at low redshifts. 
    Galaxies are labelled by how they were detected using the colour scheme in \autoref{fig:Q1431}. 
    Stars represent the sample of six possible associated galaxies.
    For this and the following figures, diamonds represent galaxies that are closest in impact parameter to the absorber (most probable host galaxies).
    }
    \label{fig:vel_b_dist}
\end{figure}

While the majority of the absorbers in the MUSE-ALMA Halos survey are associated with more than one galaxy, connecting the absorber to a `most probable host galaxy' is useful when studying the distribution of gas in the CGM.
Traditionally, the galaxy closest in impact parameter to the absorber has been designated as the most probable host \citep{Schroetter2016}, and we also choose to adopt this method. 
The distribution in impact parameter, $b$, and velocity space, $\Delta v$, of the associated galaxies is depicted in \autoref{fig:vel_b_dist}. 
The diamonds represent galaxies that are closest in impact parameter to the absorber and for two-thirds of the absorbers with associated galaxies (18/27), these galaxies also have the smallest velocity separation to the absorber.
We also find that the distribution in $b$ is not uniform, with the majority of galaxies found $< 100$ kpc from the absorber. 
This effect is partly caused by the wide range of absorber redshifts ($0.20 \lesssim z \lesssim 1.4$) corresponding to physical scales of $100$ to $250$ kpc for an angular separation of 30 arcseconds. 
The distribution of velocity differences between absorber and galaxy is found to be symmetric about $\Delta v = 0$ \kms with a tail at large positive $\Delta v$. 
To test whether the tail is statistically significant, we split the sample into two using the sign of $\Delta v$. 
Then, we take the absolute value of the sample with $\Delta v < 0$ and perform a two-sided Kolmogorov-Smirnov test to determine whether the two partitions are drawn from the same population. 
We find a $p$-value of approximately 0.3, suggesting that the $\Delta v$ values of our associated galaxies are symmetrically distributed about 0. 

\begin{figure*}
    \includegraphics[width=\textwidth]{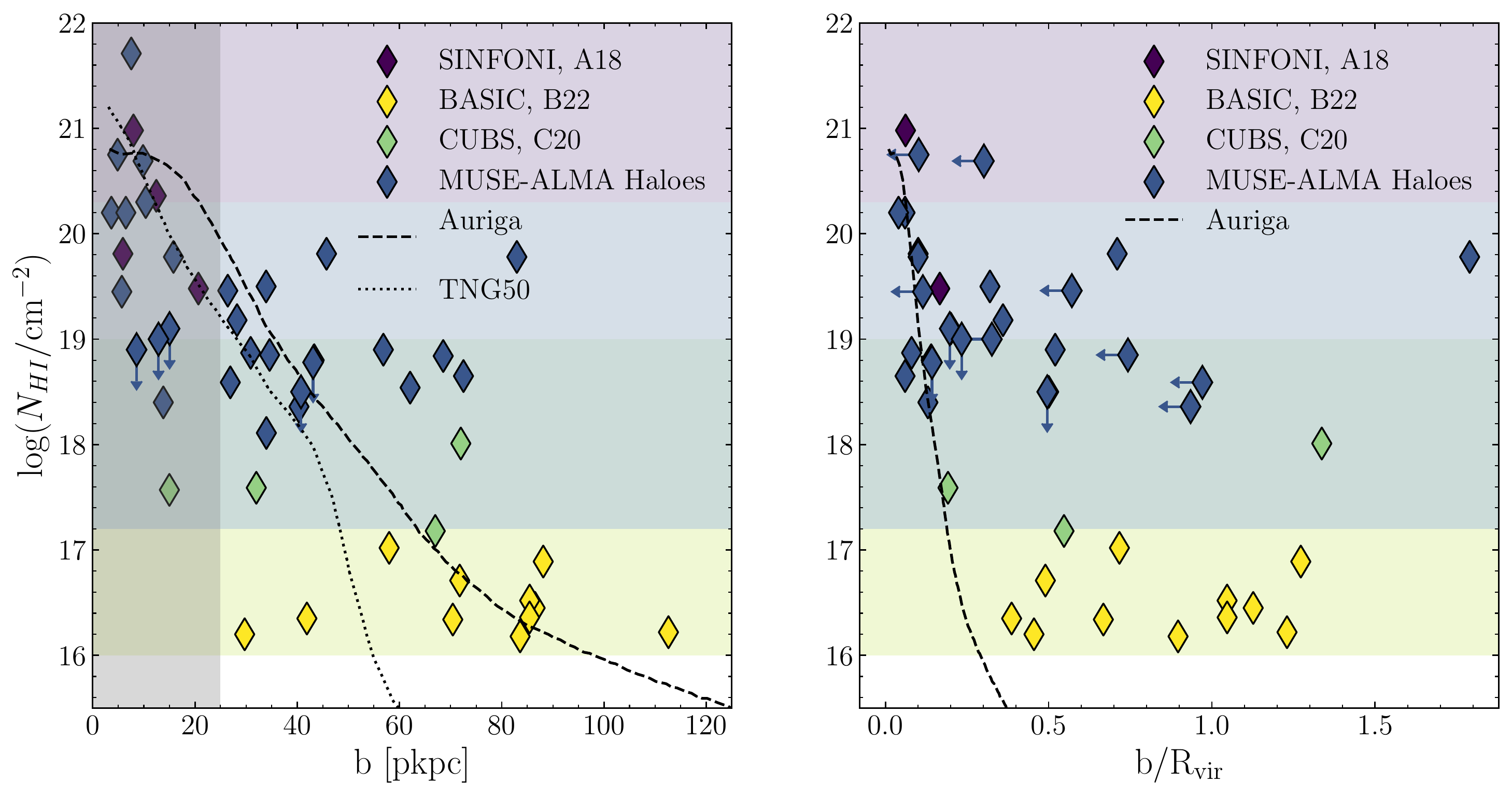}
    \caption{The \ion{H}{i} column density of absorbers against the impact parameters of associated galaxies. 
    The galaxies closest in impact parameter to the absorber are plotted here as diamonds. 
    We include data from SINFONI surveys of galaxy counterparts to absorbers \citep[A18;][]{Augustin2018}, the Bimodal Absorption System Imaging Campaign \citep[B22;][]{Berg2022} and the Cosmic Ultraviolet Baryons Survey \citep[C20;][]{Chen2020} to extend our results to higher and lower column densities. 
    The coloured regions from top to bottom represent damped Ly-$\alpha$ (DLA), sub-DLA, LLS and pseudo-LLS column densities, while the grey region marks the inner $25$ kpc where ISM gas is likely responsible for the absorption.
    The median radial profiles for the neutral hydrogen column density from the simulations by \citet{vandeVoort2019} and \citet{Nelson2020} are respectively plotted as dashed and dotted lines.
    In the right plot, we normalise the impact parameter by the virial radius. 
    For galaxies without stellar mass measurements, we derive an upper limit for the stellar mass using the SFR limit for each field and assuming the galaxy lies on the SFR-$M_*$ main sequence \citep{Schreiber2015}. 
    Passive galaxies will shift the limits leftward towards smaller values of $b$/$R_{\rm vir}$ (see Appendix \ref{App:calc} for the calculation).
    Both plots reveal a scatter of $\sim$3 dex in $N_{\ion{H}{i}}$ at a given impact parameter, but also a clear decrease in \logNHI\ at higher $b$.
    }
    \label{fig:minb_NHI}
\end{figure*}

In the left panel of \autoref{fig:minb_NHI}, we plot \logNHI \ against $b$ for galaxies closest in impact parameter ($\min(b)$) to the absorber. 
We apply this criteria to the BASIC \citep{Berg2022} and CUBS \citep{Chen2020} surveys and extend our results to lower column densities. 
Additionally, we include detections from early IFS studies that use the Spectrograph for INtegral Field Observations in the Near Infrared (SINFONI) instrument to search for galaxy counterparts to high column density absorbers \citep[e.g.][]{Bouche2007, Bouche2013, Schroetter2015, Peroux2016}.
Our $N_{\ion{H}{i}}(b)$ results when combined with previous surveys enable a comparison with simulations of the circumgalctic medium.
\citet{vandeVoort2019} uses zoom-in magnetohydrodynamic (MHD) simulations of a Milky Way-mass galaxy from the Auriga project \citep{Grand2017} and performs an additional uniform spatial refinement to enhance the physical scales resolved in the CGM. 
In \autoref{fig:minb_NHI}, we plot the median radial profile of the neutral hydrogen column density for the highest resolution simulation ($\approx$1 kpc spatial resolution of the CGM) using a dashed line.
Similarly, \citet{Nelson2020} study the distribution of cold gas in the CGM around massive galaxy group halos ($\sim\! 10^{13.5}$ M$_\odot$) at $z \sim 0.5$ by post-processing TNG50, the lowest volume and highest resolution simulation from the Illustris-TNG suite. 
Within a third of the virial radius, sub-kiloparsec resolutions are reached and this increases to $\approx 3$ pkpc at $R_{\rm vir}$. 
We choose to include both these simulations because the systems probed by \ion{H}{i} absorbers have been found to be associated with isolated galaxies and galaxy groups. 
While there appears to be agreement between the IFS observations and the median radial profile obtained from these hydrodynamical simulations out to $\sim$ 120 pkpc, we note that the \ion{H}{i} versus $b$ relation depends on the galaxy stellar mass. 
The median stellar mass $\log M_{*}$ = 9.8 of the MUSE-ALMA Halos sample is lower than a Milky Way-mass galaxy. 

We emphasise that different criteria can be used to select the most probable host galaxy of the \ion{H}{i} gas.
In Appendix \ref{app:minbv_NHI}, we reproduce \autoref{fig:minb_NHI} by forcing the associated galaxy to be closest in both impact parameter ($\min(b)$) and velocity ($\min(|\Delta v|)$) to the absorber. 
This reduces the size of the MUSE-ALMA Halos sample (18/27 associated galaxies remain), but increases the probability that the absorber is tied to the most probable host. 
The application of these criteria reduce the scatter in \logNHI]\ at a given impact parameter. 
However, the fact that numerous galaxies are found associated with these absorbers suggests gas in the CGM is not connected to the single halo of an isolated galaxy.
In the zoom-in simulation of the CGM for a Milky Way-mass galaxy, the 1$\sigma$ scatter in column density is $> 3$ dex at $b = 50$ kpc \citep{vandeVoort2019}.
For larger halo masses ($10^{13.2}-10^{13.8}$ M$_\odot$) at $z \sim 0.5$, \citet{Nelson2020} finds column densities ranging from \logNHIunit $= 14$ to $20$ at an impact parameter of 50 pkpc within a 1$\sigma$ scatter band.
The significant observed scatter in the \ion{H}{i} column density at any given $b$ appears to be an intrinsic property of the CGM that is also found in simulations and this will be discussed further in Section \ref{disc:gas_distribution}.

As the stellar masses of associated galaxies vary significantly in the sample, a more physical representation of the impact parameter can be obtained by normalizing it with respect to the virial radius and this is shown in the right panel of \autoref{fig:minb_NHI}.
We note that the stellar masses are derived using spectral energy distribution (SED) fitting of the HST and MUSE data, and this will be published in another work (Augustin et al., in prep).
We include also upper limits for galaxies where no SED fitting to determine the stellar mass could be performed.
The upper limits are derived from the SFR limit for each field assuming the galaxy is on the SFR-$M_*$ main sequence \citep{Schreiber2015}. 
While we find the majority of associated galaxies lie on the main sequence (Augustin et al., in prep), passive galaxies will shift the limits leftward towards smaller values of $b/R_{\rm vir}$. 
This is because our estimates of galaxy virial mass use a stellar-to-halo mass relation and passive galaxies will have a larger stellar mass for a given star-formation rate (see Appendix \ref{App:calc} for the calculation).
The results reveal a drop in \logNHI at $b/R_{\rm vir} > 0.5$, but also variations up to $\sim$3 dex in \ion{H}{i} column density at a given impact parameter. 
When compared with a simulated Milky Way-mass galaxy with virial radius 337 kpc \citep{vandeVoort2019}, the decline in \ion{H}{i} column density is not as precipitous.

\begin{figure}
    \includegraphics[width=\columnwidth]{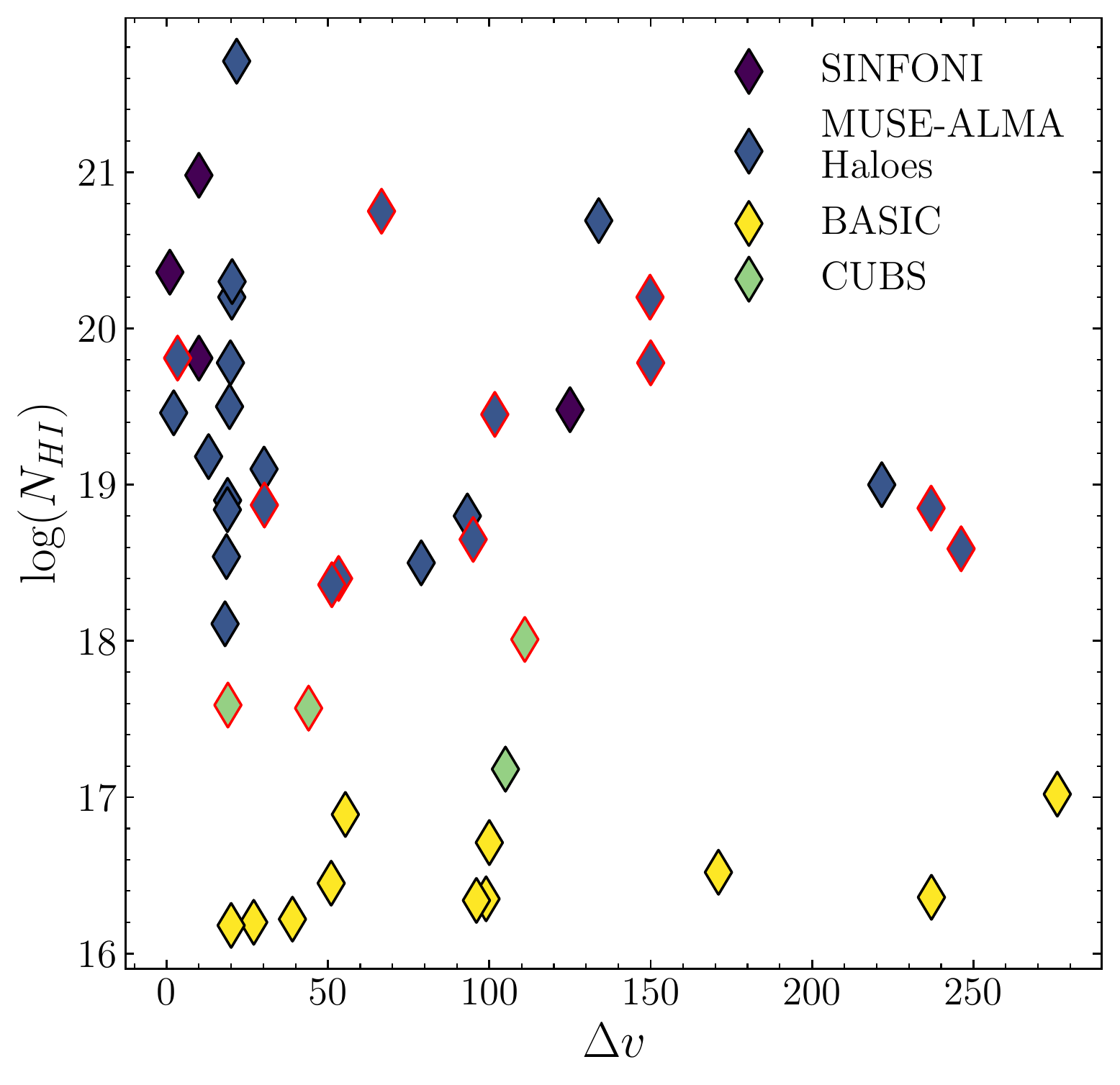}
    \caption{The \ion{H}{i} column density of absorbers against the velocity difference of associated galaxies.
    Only the most probable hosts are plotted.
    Diamonds with a red border indicate the galaxy is found at negative velocities compared to the absorber redshift.
    For objects from the BASIC survey, the tabulated velocity differences in \citet{Berg2022} did not have a sign.
    We find that the galaxies associated with high column density absorbers are not necessarily close (within 50 \kms\!) to the absorber redshift.
    }
    \label{fig:dv_NHI}
\end{figure}

We find little correlation between the absorber \ion{H}{i} column density and the velocity difference between $z_{\rm abs}$ and the galaxy at the lowest impact parameter in \autoref{fig:dv_NHI}. 
At $|\Delta v| < 50$ \kms\!, the column density spans more than five orders of magnitude. 
This is unsurprising because the variety in the origin of absorbers means that the velocity difference should not relate to the column density. 
Previous works from the MUSE-ALMA Halos survey already find the absorber attributed to large rotating disks with $V_{\rm max} = 200$ \kms \citep[$\Delta v \sim 50$ \kms\!;][]{Peroux2017}, cold-flow accretion \citep[$\Delta v \sim 80$ \kms\!;][]{Rahmani2018a} and outflowing gas \citep[$\Delta v \sim 110$ \kms\!;][]{Rahmani2018b}.

\subsection{Galaxy Star-formation Rates}
We calculate the star-formation rates (SFRs) of associated galaxies using the H$\alpha$ emission line when available. 
For sources at redshift $z \gtrsim 0.4$ where H$\alpha$ is not observable with MUSE, we estimate the SFR using the [\ion{O}{ii}] luminosity. 
There is a median discrepancy of $\sim$40 per cent in the SFRs calculated using H$\alpha$ and [\ion{O}{ii}] emission-line fluxes for the 20 associated galaxies with both lines detected.
The prescriptions from \citet{Kennicutt1998} are used to determine the SFR to keep our results consistent with past MUSE-ALMA Halos papers \citep{Hamanowicz2020}.
For galaxies with dust corrections available, we also provide a dust-corrected SFR. 
$3\sigma$ SFR limits are calculated for non-detections assuming the FWHM of the line profile is $3$ \AA\ and the emission is found in the number of pixels contained within a disk with diameter equal to the seeing. 
For each galaxy, we categorise them as star-forming if emission lines are present in the spectrum or passive if they are not. 
More rigorous criteria based on our SFR calculations are difficult to apply because the majority of the associated galaxies do not have extinction measurements. 
These values are tabulated in \autoref{tab:associated}.

\begin{figure*}
    \includegraphics[width=\textwidth]{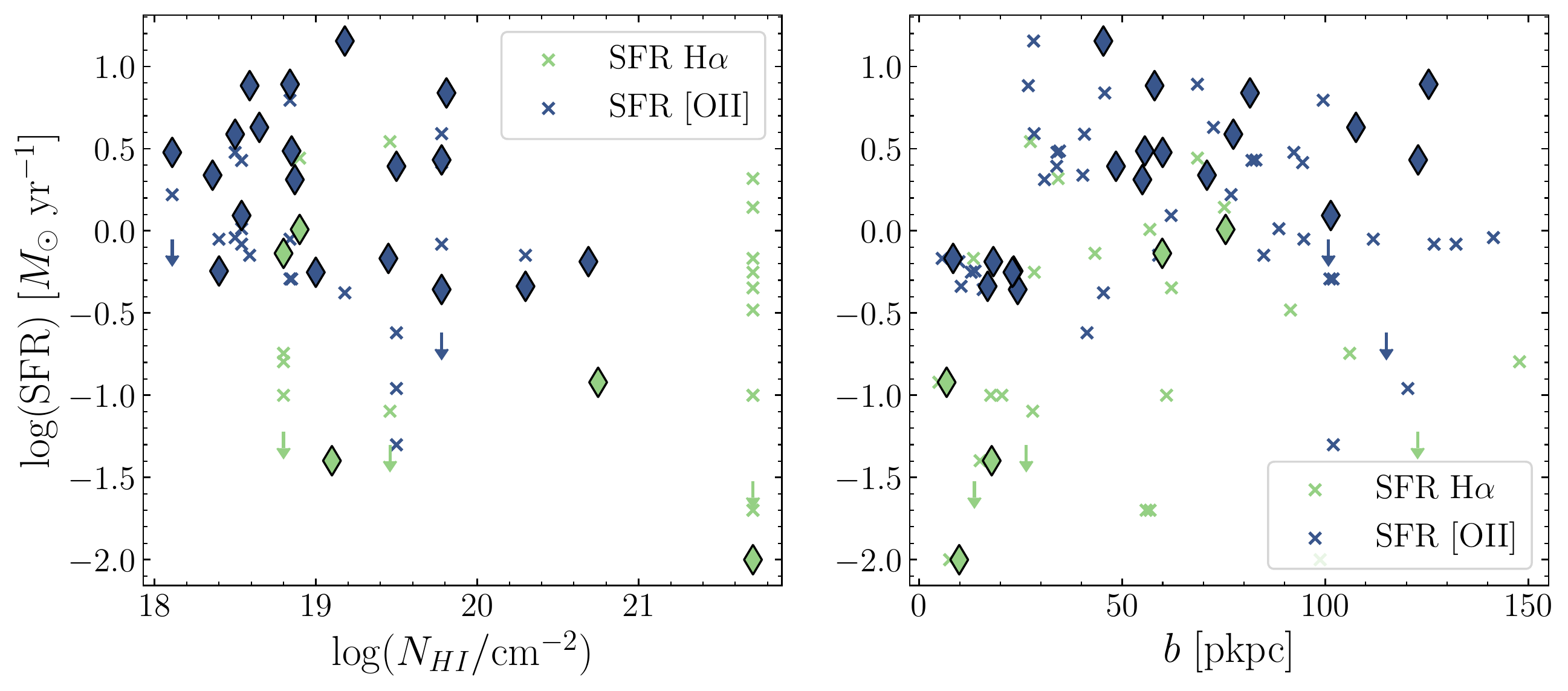}
    \caption{The star-formation rates of associated galaxies plotted against \ion{H}{i} column density (left) and impact parameter (right). 
    The SFR values use the H$\alpha$ and [\ion{O}{ii}] luminosity calibrations from \citet{Kennicutt1998}. 
    Associated galaxies closest in impact parameter to the QSO sightline are plotted as diamonds. 
    }
    \label{fig:SFR_b_NHI}
\end{figure*}

From \autoref{fig:SFR_b_NHI}, we find the star-formation rates of associated galaxies span $> 2$ dex. 
We find that the most probable galaxy hosts typically have larger SFRs than other members in their group. 
This trend is clearer in the right panel of \autoref{fig:SFR_b_NHI}, where the most probable host galaxies at larger impact parameters appear to have higher SFRs. 
However, the lack of dust corrections for $\sim$70 per cent of our sample inhibits any interpretation of these results.

\subsection{Galaxy Emission-line Metallicity}
The extensive redshift range of the absorbers ($0.2 \lesssim z \lesssim 1.40$) means different emission lines are seen for galaxies at different redshifts. 
Additionally, only galaxies at redshift $z < 0.4$ can be corrected for dust using the H$\alpha$/H$\beta$ Balmer decrement.
We choose to use the ${R}_3 = \log( F([\ion{O}{iii}] \lambdaup 5007)/ F(\rm{H}\beta))$ auroral strong line calibration from \citet{Curti2017, Curti2020a} to calculate galaxy metallicities.
This has the advantage of covering the required emission lines [\ion{O}{iii}] and $\rm{H}\beta$ up to $z \sim 0.85$ and being largely unaffected by dust obscuration. 
For objects below redshift $0.4$, we also measure ${\rm O_3  N_2} = \log( F([\ion{O}{iii}] \lambdaup 5007)/ F({\rm H}\beta))/\log(F([\ion{N}{ii}] \lambdaup 6584)/ F({\rm H}\alpha))$ \citep{Pettini2004} to obtain a check of our $\rm R_3$ measurements and in some cases, break the degeneracy in the double-branched $\rm R_3$ indicator. 
Upper and lower limits are given for objects without detectable flux in the [\ion{O}{iii}] and $\rm{H}\beta$ emission lines. 

Metallicity errors are calculated by first determining $\Delta \rm R_3$ and $\Delta \rm O_3 \rm N_2$ by propagating the flux measurement errors.
We then use Monte Carlo sampling to determine the errors in the roots of the polynomial that we solve to calculate the emission metallicity. 
We report our final metallicity values in \autoref{tab:associated}. 
For consistency, only metallicities calculated using the $\rm R_3$ calibration are reported, but we note that for galaxies with both $\rm R_3$ and ${\rm O_3  N_2}$ measurements, the metallicity values agree within their $1\sigma$ errors.

\begin{figure*}
    \includegraphics[width=0.9\textwidth]{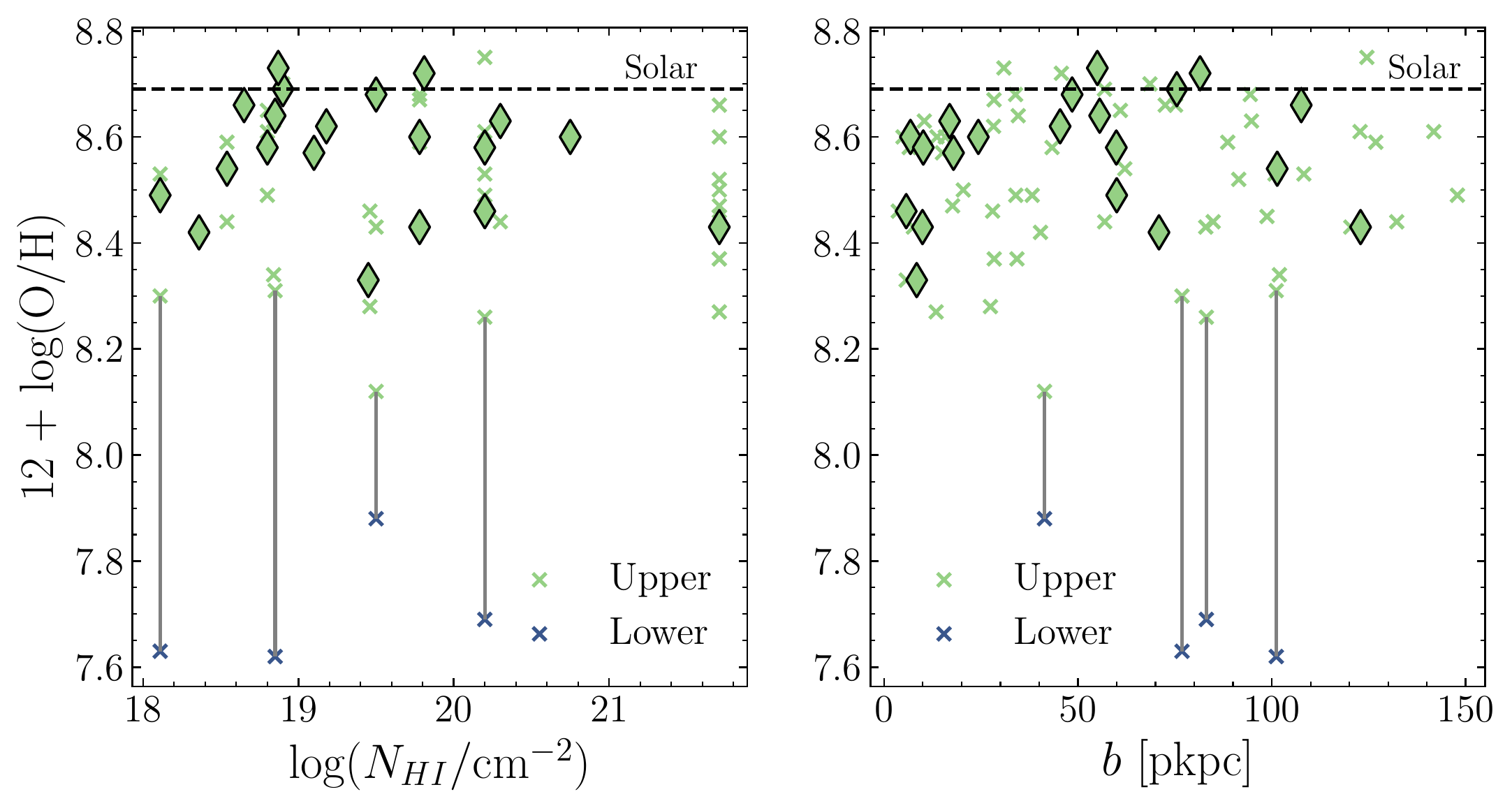}
    \caption{We plot the upper and lower branches of the associated galaxy sample using the $R_{\rm 3}$ metallicity calibration. 
    If no $\rm O_3 N_2$ measurement is available to break the degeneracy, a grey line is used to connect the upper- and lower-branch measurements. 
    The dashed line indicates the solar metallicity $(12 + \log({\rm O/H})) = 8.69$ from \citet{Asplund2009}.
    Like in previous plots, the most probable host galaxies are denoted by diamonds. 
    The left and right panels reveal no trend between the metallicity of the associated galaxy and \ion{H}{i} column density or impact parameter respectively. 
    }
    \label{fig:Z_b_NHI}
\end{figure*}

We find no significant trend between galaxy metallicity and absorber column density in the left panel of \autoref{fig:Z_b_NHI}. 
However, we find that the galaxy metallicities are typically sub-solar \citep[assuming $(12 + \log({\rm O/H})) = 8.69$;][]{Asplund2009}. 
The right panel of the same figure plots galaxy metallicity against impact parameter and again appears largely scattered.

Roughly half the absorbers in the sample (17/32) have Zn or Fe abundance measurements or limits from the literature. 
We preferentially adopt the [Zn/H] metallicity when available because Zn is typically not significantly depleted onto dust in the ISM \citep{Spitzer1975, Savage1996, Jenkins2009, Vladilo2011}.
Otherwise, we correct our [Fe/H] measurements for dust by adding 0.3 to the original [Fe/H] value \citep{Rafelski2012}.
Seven absorbers with metallicity limits have $\logNHIunit < 19.5$ where an ionization correction might be applicable.

\begin{figure}
    \includegraphics[width=\columnwidth]{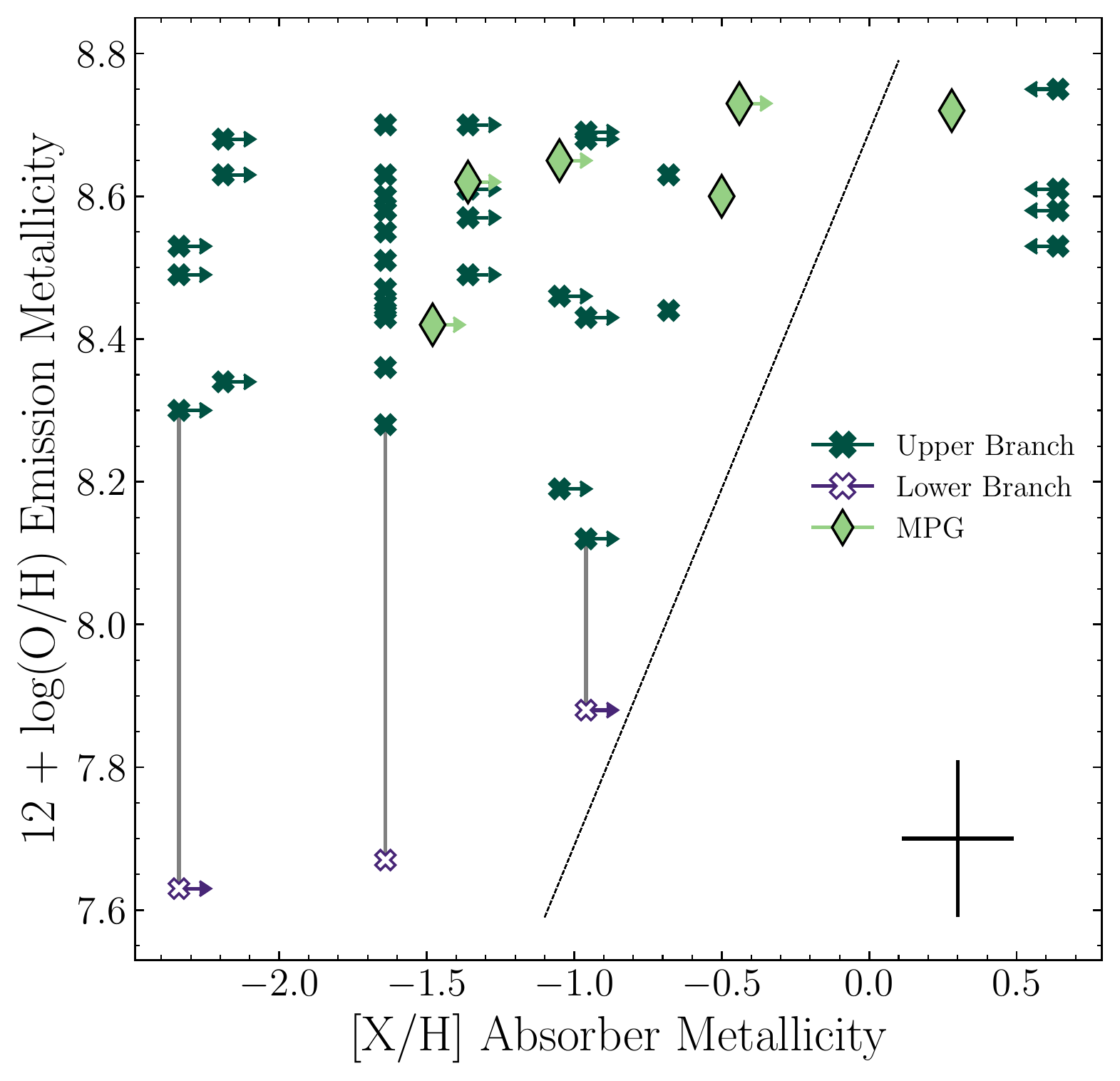}
    \caption{The absorber metallicity compared with the galaxy emission-line metallicity. 
    The most probable galaxy hosts of the \ion{H}{i} absorbers are denoted by the diamonds, while the remaining galaxies are crosses. 
    Vertical lines connect the upper and lower branch galaxy metallicities that are degenerate using the $\rm R_3$ calibration.
    17 out of the 32 absorbers have metallicity measurements [Zn/H] or [Fe/H] and 10 of these measurements are limits. 
    The dotted line denotes the line of equal galaxy and absorber metallicity. 
    We find that the metallicity measured from strong-line diagnostics is typically larger than the CGM absorbing gas.
    }
    \label{fig:Z_abs_gal}
\end{figure}

We plot the galaxy emission metallicities found using the $\rm R_3$ calibration against the absorber metallicity. 
\autoref{fig:Z_abs_gal} reveals that the emission-line metallicities are higher than the metallicities probed by absorbers in the CGM of galaxies. 
This finding broadly agrees with previous comparisons between galaxy and CGM metallicities where a gradient is found, but the scatter is large \citep{Chen2005, Christensen2014, Peroux2014}. 

\begin{figure}
    \includegraphics[width=\columnwidth]{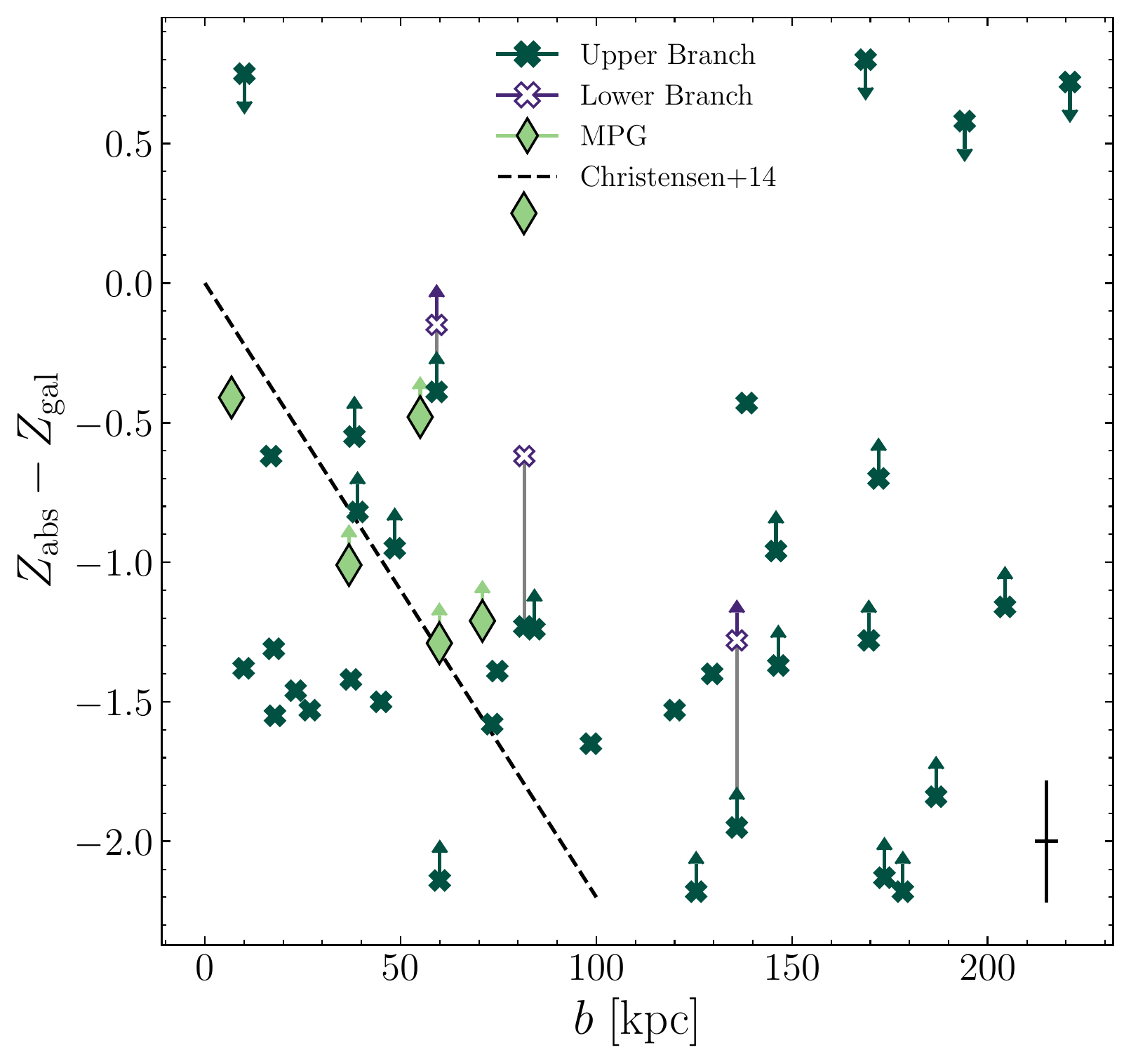}
    \caption{The difference in absorber and galaxy metallicity plotted against impact parameter. 
    The most probable galaxy hosts of the \ion{H}{i} are denoted by the diamonds, while the remaining galaxies are crosses. 
    We include results from previous works studying damped Lyman-$\alpha$ absorbers and their host galaxies \citep{Peroux2013, Christensen2014}.
    Vertical lines connect the upper and lower branches galaxy metallicities that are degenerate using the $\rm R_3$ calibration.
    The measured dependence of metallicity on impact parameter from \citet{Christensen2014} is included as a dashed black line. 
    }
    \label{fig:deltaZ_abs_gal}
\end{figure}

The comparison between galaxy and absorber metallicity requires careful interpretation. 
Emission-line diagnostics such as $\rm R_3$ and $\rm O_3N_2$ measure the integrated oxygen abundance in a galaxy. 
On the other hand, the absorption metallicity is a measure of the Zn or Fe abundance in a pencil-beam sightline intersecting gas in the CGM at some impact parameter from the galaxy centre. 
The correlation between this emission metallicity and the Zn or Fe abundance found in gas clouds is unclear. 
However, we find that the majority of galaxies have \ion{H}{ii} region metallicities in their inner regions that are much higher than those measured in absorption in their outskirts. 
These differences are significantly larger than those expected from an intrinsic enhancement in the ratio [O/Fe] compared to solar. 
Thus, the offsets we see in \autoref{fig:deltaZ_abs_gal}, where we show the metallicity difference between the absorber and the galaxy as a function of impact parameter, are caused by more than just a mismatch in the elements used to deduce the metallicity.

For the galaxy associated with the absorber towards Q0138$-$0005 at $z_{\rm abs} = 0.7821$, we find the absorption metallicity to be larger than the emission-line metallicity of the galaxy (top-right diamond in \autoref{fig:Z_abs_gal}). 
This is unusual because we expect the absorber abundance measurement to be significantly lower than the mean metallicity of the galaxy at an impact parameter of $81$ kpc \citep{Christensen2014}. 
However, for this field, we reach a limiting magnitude limit of only $m_r = 23.5$ and it is possible that the QSO sightline intersects a fainter, undetected galaxy at smaller impact parameters. 
Further discussion on the non-detections of galaxy counterparts to the five absorbers can be found in Section \ref{disc:counts}.

\subsection{[\ion{O}{iii}]/H$\beta$ versus [\ion{O}{ii}]/H$\beta$ Diagram}
For absorbers with $z_{\rm abs} \gtrsim 0.4$, the absence of H$\alpha$ and [\ion{N}{ii}] means the classical Baldwin, Phillips \& Terlevich (BPT) diagram \citep{BPT1981} cannot be used to distinguish between star-forming galaxies and active galactic nuclei. 
Instead, we use the blue classification diagram from \citet{Lamareille2010} to separate our sample into star-forming, Seyfert 2 and low-ionisation nuclear emission-line region (LINER) galaxies. 
Equivalent width (EW) ratios of [\ion{O}{iii}] $\lambdaup 5007$/H$\beta$ and [\ion{O}{ii}]$\lambdaup \lambdaup 3727,29$/H$\beta$ are computed for all associated galaxies.
We exclude galaxies with continuum levels below $5 \times 10^{-19}$ erg s$^{-1}$ cm$^{-2}$ \AA \ adjacent to the relevant lines to ensure our EW measurements are accurate.
Additionally, we limit the sample to galaxies with SNR $> 3$ in [\ion{O}{ii}], H$\beta$ and [\ion{O}{iii}] emission. 
The wavelength range of MUSE also limits the redshifts of galaxies in this sample to $0.26 \lesssim z \lesssim 0.85$.

\begin{figure}
    \includegraphics[width=\columnwidth]{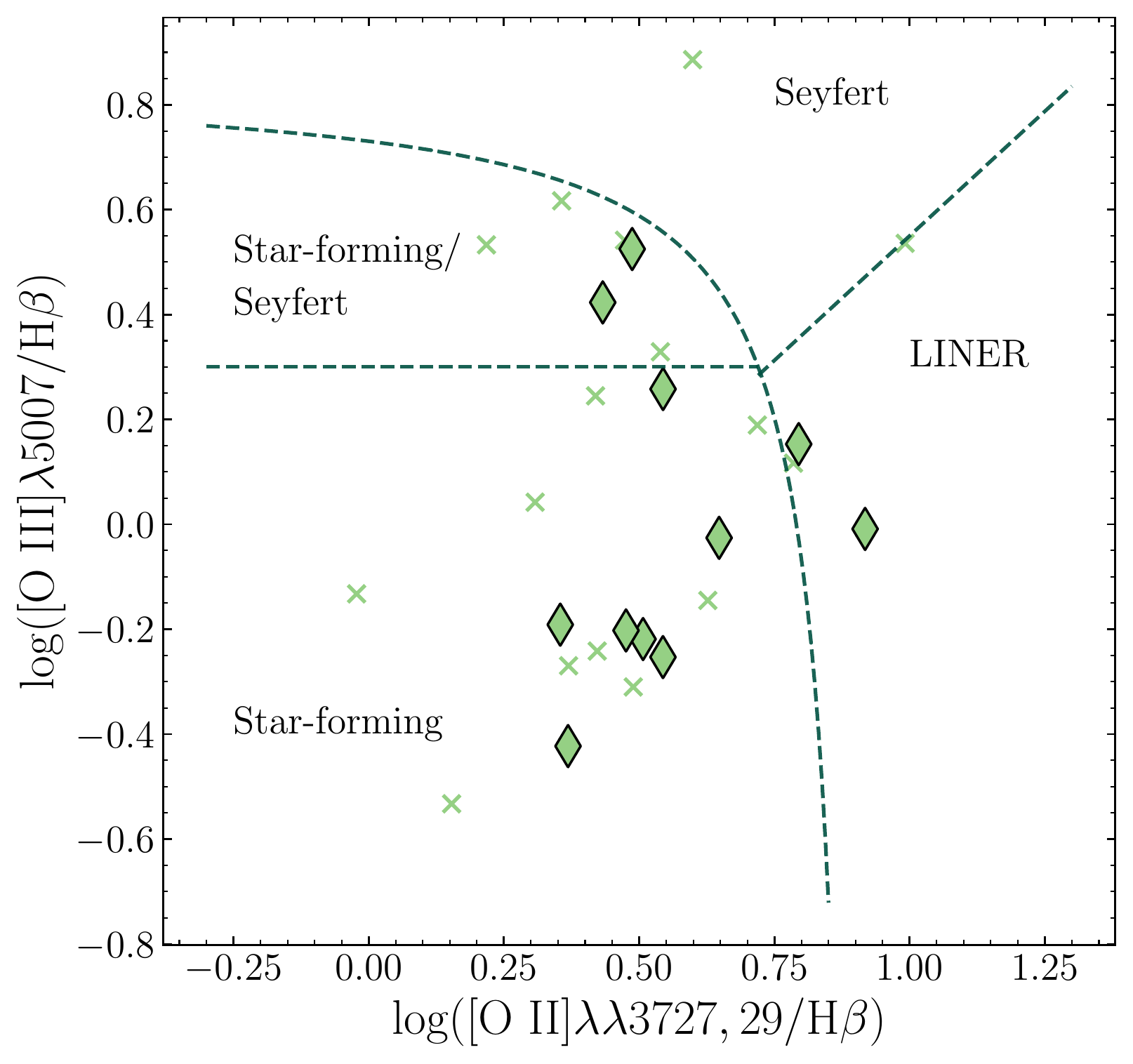}
    \caption{The [\ion{O}{iii}]/H$\beta$ versus [\ion{O}{ii}]/H$\beta$ classification diagram for galaxies associated with \ion{H}{i} absorbers. 
    The dashed line represents the empirical separation between star-forming galaxies, Seyfert 2 galaxies and LINERs.
    }
    \label{fig:BPT}
\end{figure}

In \autoref{fig:BPT}, we plot the 27 associated galaxies that meet the aforementioned criteria on the classification diagram.
We find that 16/27 galaxies are star-forming, with the remainder being placed in the Seyfert 2 and LINER regions of the diagram.

\section{Discussion}
The MUSE-ALMA Halos survey has uncovered 79 galaxies associated with 32 \ion{H}{i} Ly-$\alpha$ absorbers.
These absorbers vary significantly in column density ($18 \lesssim \logNHIunit \lesssim 21.7$) and cover redshift $0.19 < z < 1.15$. 
The majority (17 out of 32) have two or more galaxies within $\pm 500$ \kms of the \ion{H}{i} gas and only five absorbers do not have any associated galaxy detected in the MUSE data. 
In this section, we discuss the implications of our results on the distribution of gas and metals in the CGM and how galaxy properties such as SFR and metallicity correlate with absorber properties.

\subsection{Number counts of galaxies associated with \ion{H}{i} absorbers}
\label{disc:counts}
\subsubsection{Galaxy Overdensities}
Galaxy overdensities are increasingly being found at the redshift of absorbers.
Compiling our results with surveys such as CUBS \citep{Chen2020} and BASIC \citep{Berg2022}, we find that of the 48 total absorbers with MUSE data, $\sim$50 per cent (23/48) are associated with two or more galaxies. 
There are 8 non-detections of galaxies within the vicinity of absorbers and $30$ per cent (16/48) have only one associated galaxy. 
We note that these IFS studies vary significantly in depth ($\sim$3 magnitudes in the $r$-band), and it is likely that the proportion of \ion{H}{i} absorbers associated with two or more galaxies is higher in reality than indicated by the combined results of these surveys. 
Moreover, the number of associated galaxies also depends on the selection criteria.
For the values given above, we count galaxies within $\pm 500$ \kms of the absorber to be associated but emphasise that stricter criteria will alter the results. 

For all absorbers with $> 2$ galaxies within $500$ \kms of the absorber redshift, we compute the expected number of galaxies within a co-moving volume using the [\ion{O}{ii}] and [\ion{O}{iii}] luminosity functions scaled by $z$ from \citet{Comparat2016} (the latter only when [\ion{O}{ii}] flux measurements are unavailable).
We assume a cylindrical volume with diameter equal to the largest separation between associated galaxies and height equal to the co-moving distance between the minimum and maximum associated galaxy redshifts. 
The number of galaxies found are on order $10^{1}-10^{3}$ larger compared to the expected number (see \autoref{tab:group_properties}). 
Importantly, the high $N_{\ion{H}{i}}$ values of the absorbers means that we know \textit{a priori} that there is likely a galaxy at the absorber redshift. 
Our calculations of the expected number of galaxies using a luminosity function does not take this into account.

\begin{table}
\begin{center}
\caption{\textbf{Properties of the galaxy overdensities found at the absorber redshift.} 
The number of expected galaxies for each absorber are calculated using the [\ion{O}{ii}] and [\ion{O}{iii}] luminosity functions \citep{Comparat2016}.
We also include estimations of the group dispersion, $\sigma_{\rm group}$, and virialized group mass, but note these are lower limits because there may be galaxies outside the MUSE field-of-view. 
For the $z_{\rm abs} = 0.3929$ absorber towards Q1211$+$1030, the three associated galaxies have a maximum velocity separation of $\sim$25 \kms between them which results in the smaller $N_{\rm expected}$, $\sigma_{\rm group}$ and group mass measurements.
}
\begin{tabular}{llcccc}
\hline\hline
Quasar & $z_{\rm abs}$ & $N_{\rm expected}$ & $N_{\rm gal}$ & $\sigma_{\rm group}$ & Group Mass \\
    &  &  &  & (\kms\!) & ($\log$M$_\odot$) \\
\hline 
Q0152$+$0023 & 0.4818 & 0.48 & 4  & 190 & 12.4  \\
Q0152$-$2001 & 0.3830 & 0.40 & 5  & 100 & 12.3  \\
Q0420$-$0127 & 0.6331 & 0.10 & 4  & 110 & 12.1 \\
Q1110$+$0048 & 0.5604 & 0.04 & 3  & 100 & 12.0 \\
Q1130$-$1449 & 0.3130 & 0.18 & 13 & 120 & 11.8 \\
Q1211$+$1030 & 0.3929 & 0.002& 3  & 13  & 10.0  \\
Q1211$+$1030 & 0.8999 & 2.0  & 3  & 190 & 12.9 \\
Q1229$-$021  & 0.7691 & 0.07 & 3  & 150 & 12.3 \\
Q1229$-$021  & 0.8311 & 0.28 & 4  & 67  & 12.4 \\
Q1515$+$0410 & 0.5592 & 0.07 & 4  & 22  & 11.1 \\
Q2131$-$1207 & 0.4298 & 0.12 & 4  & 83  & 12.0 \\
\hline\hline 				       			 	 
\label{tab:group_properties}
\end{tabular}			       			 	 
\end{center}			       			 	 
\end{table}	

We also estimate group masses for all absorbers with more than three associated galaxies and these values are tabulated in \autoref{tab:group_properties}. 
A \textit{gapper} function \citep{Beers1990} is used to determine the velocity dispersion $\sigma_{\rm group}$ between group members and the virial theorem is applied ($M_{\rm group} \sim 3R\sigma_{\rm group}^2/G$), assuming the radius $R$ is half the largest separation between galaxies in the field. 
We emphasise this estimate is a lower limit because many systems likely contain objects outside the MUSE field-of-view. 

\subsubsection{Absorbers without associated galaxies}
We find that for five absorbers, there are no galaxies found within 500 \kms of the absorber redshift. 
There are three possible causes for this: i) the absence of galaxy counterparts is due to an intrinsic property of the absorber such as \ion{H}{i} column density or metallicity, ii) there are galaxies outside the MUSE field-of-view, or iii) there are galaxies below our detection limits.

The absorbers with missing galaxy counterparts span column densities of $\logNHIunit =18.62$ to 21.54.
Additionally, three out of the five absorbers have measured [Zn/H] abundances of $-0.92$, $-0.45$ and $0.08$. 
\citet{Berg2022} found the population of pLLS and LLS systems to be bimodal and metal-poor absorbers (abundances [X/H] $\leq -1$) not associated with any galaxies. 
The work suggests a population of \logNHIunit $<17.0$ absorbers may trace gas associated with overdense regions of the Universe rather than galaxy halos. 
In the MUSE-ALMA Halos survey, the higher column density of the absorbers studied means the gas is more likely found in galaxy halos rather than these purported overdense regions. 
The metal abundances of the absorbers without galaxy counterparts are also enhanced compared to the values found in \citet{Berg2022}.

Another possibility is that there are galaxies outside the MUSE field-of-view. 
The lowest redshift absorber without an associated galaxy is towards Q2353$-$0028 at $z_{\rm abs} = 0.6044$ where 30 arcseconds corresponds to $> 200$ kpc. 
In the MUSE-ALMA Halos sample, $\sim$95 per cent of associated galaxies are found within 200 kpc which indicates the missing galaxy counterparts are likely not beyond the FoV. 
Expanding our search to $|\Delta v| \leq 3000$ \kms reveals only one galaxy with velocity difference $-$2685.2 \kms for the absorber towards Q2353$-$0028. 
However, the extreme column density ($\logNHIunit = 21.54$) of this system suggests the host galaxy is within a few kpc ($<$ 1 arcsecond) of the absorber \citep{Noterdaeme2014}. 
Hence, the associated galaxy may be a passive dwarf galaxy within the point spread function (PSF) of the QSO. 
In fact, there are two more absorbers (Q$0058+0019$ at $z_{\rm abs} = 0.6127$ and Q$0123-0058$ at $z_{\rm abs} = 1.4094$) with column densities $\logNHIunit > 20$ and no galaxy counterpart. 
The results in the left panel of \autoref{fig:minb_NHI} suggest the associated galaxy should be within 20 kpc of the QSO sightline and this is consistent with the characteristic sub-DLA radius \citep{Peroux2005}. 

While there are no associated galaxies detected in the MUSE data, the HST imaging after QSO subtraction reveals faint $> 25$ mag objects within several arcseconds of the quasar for the Q$0058+0019$, Q$1211+1030$ and Q$2353-0028$ fields. 
These are candidate galaxy counterparts to the absorbers but no emission lines were detected at their positions in the MUSE data. 
Thus, the absorbers without confirmed associated galaxy detections in the MUSE data may have a nearby, passive galaxy.
There is no HST imaging for the Q$0123-0058$ field so we could not search for galaxies near the QSO. 
The complete set of full and QSO-subtracted HST images will be presented in Karki et al. (in preparation). 

Recent works studying the properties of absorbers passing within $\sim$50 kpc of quiescent galaxies have found enhanced $N$(\ion{Fe}{ii})/$N$(\ion{Mg}{ii}) values compared to star-forming galaxies, suggested to arise from type 1a supernovae (SNe 1a) chemically enriching cool gas in the ISM and outer CGM \citep{Zahedy2016, Zahedy2017, Boettcher2021}. We note that here  the insufficient depth in the MUSE data (particularly the Q0058$+$0019 and Q0123$-$0058 fields) and high redshift of absorbers limits the ability to detect galaxy counterparts, without necessarily indicating the galaxies associated with the absorbers would be passive. Indeed,  \autoref{tab:abs} indicates that even star-forming galaxies will have emission fluxes below the detection limits in these data cubes. Nonetheless, we compute $N$(\ion{Fe}{ii})/$N$(\ion{Mg}{ii}) values for the absorbers for which the data are available. We find that Fe/Mg values of $-0.5$ and $-0.26$ \citep{Pettini2000, Nielsen2016} for two un-detected absorbers with measured ion column densities (Q0058$+$0019 at $z_{\rm abs} = 0.6127$ and Q1211$+$1030 at $z_{\rm abs} = 1.05$ respectively) are comparable to the values of Fe/Mg $= -0.23$ and $-0.75$ \citep{Muzahid2016, Pointon2019} for the sample of galaxies with detected associated galaxies. We will expand on this interesting point when more absorption metallicity measurements become available.



The large span in the intrinsic properties of the absorbers without galaxy counterparts, and the large MUSE FoV compared to the expected location of associated galaxies signify that there are likely galaxies below the detection limit of the MUSE observations. 
Indeed, three of the five absorbers without galaxy detections belong to the fields with the worst seeing conditions (seeing FWHM = $1.23$ for Q0058$+$0019 and $2.11$ for Q0123$-$0058). 
The magnitude limit is likewise poorer than in other fields and there are likely galaxies with SDSS $r$-band magnitudes $>24$ not detected by MUSE. 
The final absorber without a galaxy counterpart is towards QSO J1211$+$1030 at $z_{\rm abs} = 1.0496$. 
At this redshift, a main-sequence galaxy with star-formation rate $\log(\rm{SFR}) < 8.9$ \Moyr would not be detected. 
We conclude that our non-detections of associated galaxies are most likely caused by insufficient depth in the observations.


\subsection{Star-formation rates and BPT Diagram}
In the right panel of \autoref{fig:SFR_b_NHI}, we find that the star-formation rate appears to be higher for the most probable host galaxies at larger impact parameters. 
We plot in \autoref{fig:densities} the average SFR of all associated galaxies in bins of the velocity difference to the absorber (|$\Delta v$|) and impact parameter $b$. 
The bins have been artificially constructed such that there are approximately $10$ galaxies in each bin. 
There appears to be no trend in the SFR as we approach larger $|\Delta v|$ and $b$ bins. 

From the [\ion{O}{iii}]/H$\beta$ versus [\ion{O}{ii}]/H$\beta$ diagram in \autoref{fig:BPT}, we find that the majority of galaxies (16/27) fall within the star-forming region. 
The expected percentage of active galactic nuclei at the median redshift of this sample $z \sim 0.39$ in galaxies with stellar mass $\log M_{*}$ = 10.1 is $0.5$ per cent \citep{Aird2018}. 
We use the median $M_{*}$ of this sample, but note that this fraction is limited to AGN with specific black hole accretion rates $> 0.01$ (the rate of black hole accretion normalised by the galaxy stellar mass). 
The median stellar mass for this flux-limited sample of galaxies is also higher than the larger population of associated galaxies in the MUSE-ALMA Halos survey.
Hence, while we find a larger fraction of AGN in our sample of associated galaxies (20 per cent of galaxies lie in the Seyfert 2 and LINER regions of the diagram) than normal, it remains unclear whether galaxies associated with \ion{H}{i} absorbers are more likely to host AGN. 

\subsection{Distribution of gas in the CGM}
\label{disc:gas_distribution}
We see in \ion{H}{i} 21-cm emission maps of galaxies in the local Universe that the \ion{H}{i} column density decreases with radius until the ionizing radiation field causes a sharp truncation \citep{Maloney1993, Zwaan2005}. 
Extending beyond the neutral gas disk to the circumgalactic medium, similar trends have been found where \logNHI\ decreases as impact parameter increases \citep[e.g.][]{Lehner2013, Tumlinson2013, Prochaska2017}. 
The range of neutral gas column densities often spans $\sim$3 dex for a given impact parameter, although it is important to acknowledge that part of this observation may be attributed to the galaxy-to-galaxy variation within the associated galaxy sample.
Nevertheless, the scatter is significant enough that it indicates the CGM is `clumpy' and the possible reasons for this are manifold.
Recent work using high-redshift lensed quasars or galaxies as a background source show variations in the \ion{H}{i} column density of up to 1 dex in physical separations of only $2 - 9$ kpc \citep[][and references therein]{Kulkarni2019, Lopez2020, Cashman2021, Bordoloi2022}.
A factor of 5 difference in optical depths is even found on parsec scales using $21-$cm absorption towards compact symmetric objects \citep{Biggs2016}. 
Combining observations towards a quadruply lensed quasar with simulations, \citet{Augustin2021} finds fractional variations in \ion{Mg}{ii} equivalent width (EW) are dependent on the underlying structure probed by the absorber. 
While outflows are found to be `clumpy', inflowing filaments have smaller variations in EW. 
Recent gravitational-arc tomography of a galaxy at $z = 0.77$ also finds large variations in \ion{Mg}{ii} equivalent width in a galaxy's extended rotating disk \citep{Tejos2021}.
These studies reveal the CGM may be intrinsically clumpy, leading to the broad range of column densities observed at all impact parameters. 

Moreover, $\sim$50 per cent of the absorbers are associated with more than one galaxy. 
While our criteria likely select the host of the \ion{H}{i} absorber, they do not account for situations where the gas is tidally stripped or part of the intragroup medium (IGrM). 
The view of a one-to-one correlation between the absorber and the associated galaxy may contribute to the scatter seen in the relationship between \logNHI\ and impact parameter. 
It is clear that neutral gas in the CGM does not follow a homogeneous and uniformly decreasing distribution out to larger radii.  
Despite this, our observations still indicate that lower column density gas is found further away from the galaxy centre, extending the trend seen in \ion{H}{i} $21-$cm maps of local galaxies \citep[e.g.][]{Elagali2019, Kleiner2019} to the circumgalactic medium. 

\begin{figure*}
    \includegraphics[width=\textwidth]{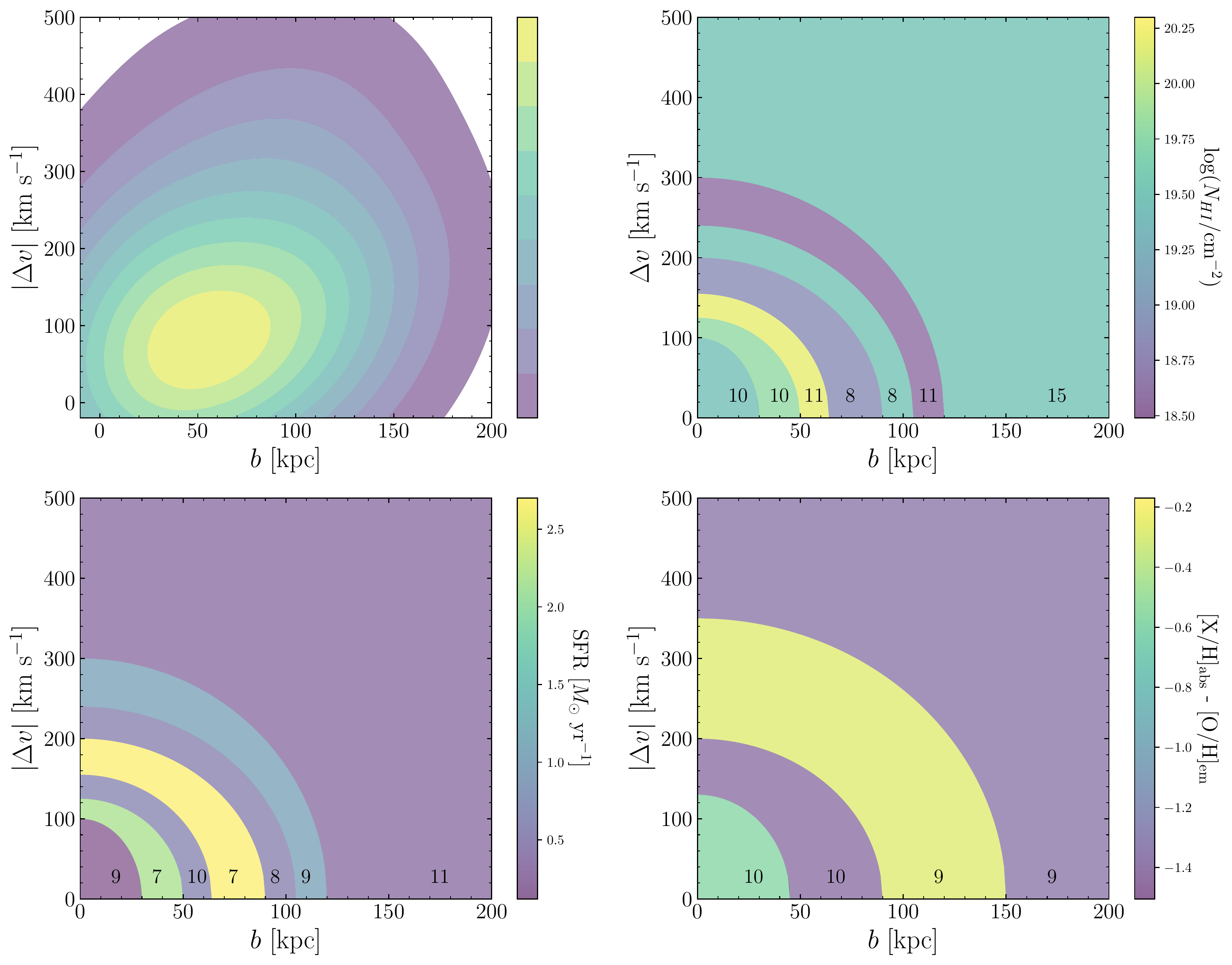}
    \caption{Median properties of galaxies and absorbers from the MUSE-ALMA Halos survey at varying impact parameters and velocities. 
    The distribution of gas and metals in the CGM can be mapped by a sufficiently large sample of QSO-galaxy pairs.
    The radial bins are distributed such that there are $\sim$10 galaxies in each bin (see labels within each bin for the exact value). 
    Velocity and impact parameter in the plots extents are artificial and caused by forcing associated galaxies to be within $\pm 500$ \kms\!.
    Each bin is coloured by the median value of the property being plotted.
    Top left: an absorber-centric smoothed kernel density estimation of galaxy number density with respect to the absorber. 
    The majority of galaxies are located within $100$ kpc and $100$ \kms of the absorber (warmer colours represent regions of higher density).
    Top right: a galaxy-centric view of the median absorber \ion{H}{i} column density at varying $|\Delta v|$ and $b$. 
    We find a decrease in the column density towards higher velocities and impact parameters.
    Bottom left: the median SFR of galaxies as a function of $|\Delta v|$ and $b$.
    Galaxies near the absorber appear to have larger SFRs than those located further away, but the trend is not consistent across the bins.
    Bottom right: the median emission-line metallicity of galaxies as a function of $|\Delta v|$ and $b$. 
    There appears to be no trend in Z$_{\rm gal}$ with increasing velocity difference or impact parameter. 
    }
    \label{fig:densities}
\end{figure*}

We plot in the top-right of \autoref{fig:densities} the median column density of absorbers found at different velocities and impact parameters from the associated galaxy. 
The plot can be interpreted as being `galaxy-centric'; the galaxy is at the origin and absorbers are placed afield.
We find that galaxies associated with lower column density systems are typically found at larger velocities and impact parameters to the absorber. 
To test the impact of bin selection on this trend, we choose varying bin sizes in \autoref{fig:HI_wQ1130}. 
Additionally, we also exclude the 13 galaxies associated with the $z_{\rm abs} = 0.313$ absorber towards Q1130$-$1449 to test whether removing this large galaxy group has a measurable impact on our results (\autoref{fig:HI_noQ1130}). 
For these cases, we still find that the \ion{H}{i} column density decreases at larger $|\Delta v|$ and $b$ values. 

While a clear anti-correlation between \logNHI\ and impact parameter is seen in \autoref{fig:minb_NHI}, there is no equivalent trend when plotting the column density against the velocity difference (\autoref{fig:dv_NHI}).
Thus, it becomes important to check whether the change in \ion{H}{i} column density is primarily dictated by the impact parameter. 
We randomly shuffle the velocity differences between each associated galaxy and absorber, and find $|\Delta v|$ plays little role in the changing column density (see \autoref{fig:HI_randomV}).
In contrast, randomly shuffling $b$ produces a plot with little variation in \ion{H}{i} column density across the bins (\autoref{fig:HI_randomb}).

Finally, we respectively normalise both $|\Delta v|$ and impact parameter by the escape velocity and virial radius of each associated galaxy in \autoref{fig:HI_norm}. 
Details on this calculation are provided in Appendix \ref{App:calc}.
Again, the column density decreases at larger virial radii, but the trend continues until $|\Delta v|/v_{\rm esc} \approx 1$ and $b/R_{\rm vir} \approx 3$. 
This suggests that neutral gas at velocities beyond the escape speed and distances several times the virial radius may no longer be associated with a galaxy.
Overall, these results suggest that while the CGM may be inhomogeneous on scales less than several kpc, a statistical study can still reveal the broader change in the column density out to larger radii. 

\subsection{Distribution of metals in the CGM}
Metallicity gradients can be found in studies of galaxy interstellar media in the local Universe \citep[e.g.][]{Sanchez2014, Belfiore2017} and at higher redshift \citep[e.g.][]{Carton2018, Curti2020b}. 
The traditional inside-out growth model of galaxy formation suggests metallicity gradients will be negative, that is, the inner parts of the galaxy are more chemically enriched than the outer parts \citep{Portinari1999}. 
However, physical processes such as outflows, inflows and mergers will affect the gradient, flattening it out \citep{Kewley2010, Bresolin2012} or even inverting it \citep{Sanchez2018}.
Further, moving from this `inner' metallicity gradient to a gradient in the CGM presents a further challenge \citep[e.g.][]{Peroux2012, Christensen2014, Peroux2016, Rhodin2018, Kacprzak2019, Kulkarni2019}.

Simulations from both EAGLE and TNG50 predict that despite the complexity of the physical processes of gas recycling (accretion, ejection, reaccretion) and multiple origin along the los, the net metallicity of CGM gas is found to be higher along the minor axis because of outflows \citep{Peroux2020}.
\citet{Freeke2021} reach the same result using Auriga zoom-in simulations of a Milky Way-mass galaxy but note that the introduction of magnetic fields decreases the mixing efficiency of inflowing and outflowing gas and subsequently increases the scatter in metallicities seen at a given azimuthal angle. 
While recent work from the MEGAFLOW survey suggests that gas along the minor axis is more dust depleted and metal enriched than the gas along the major axis \citep{Wendt2021}, a larger sample is required to observationally constrain the relationship between metallicity and azimuthal angle and account for the inherent inhomogeneity of the CGM \citep{Zahedy2019, Zahedy2021, Bordoloi2022, Sameer2022}. 
Despite these effects causing the metallicity to vary on small scales, predictions of a global trend in metallicity versus azimuthal angle are yet to be fully tested.

In \autoref{fig:Z_abs_gal}, We find that $> 90$ per cent of galaxies have confirmed emission-line metallicities larger than that of the absorbers. 
\autoref{fig:deltaZ_abs_gal} reveals that the difference between the gas-phase metallicity measured via absorption and emission lines ([X/H]$_{\rm abs} -$ [O/H]$_{\rm em}$) appears to increase in magnitude at higher impact parameters. 
However, the data required to constrain the relationship is currently too limited at $b > 40$ kpc to test whether the negative gradient found in previous studies extends to larger impact parameters \citep{Chen2005, Christensen2014, Peroux2014}.
In the bottom-right panel of \autoref{fig:densities}, we colour the radial bins by the median value of [X/H]$_{\rm abs} -$ [O/H]$_{\rm em}$. 
We also include the sample of galaxies from the BASIC survey \citep{Berg2022} with available galaxy and absorber metallicities. 
There appears to be a decrease in the metallicity from the first two radial bins (same trend seen with different binning in Appendix ref). 
Beyond the first two bins, the metallicity difference between absorber and galaxy begins to fluctuate. 
Part of the explanation for this is the fact that gas at these larger |$\Delta v$| and $b$ are beyond the virial radius of the galaxy and we expect the trend to no longer persist. 
More significantly, one must be cautious when comparing [X/H]$_{\rm abs}$ with [O/H]$_{\rm em}$. 
The former is a Zn or Fe abundance of the neutral phase in a gas cloud likely $\sim$100 pc in size while the oxygen abundance estimated from strong-line calibrations is an average ISM metallicity of the ionized gas. 
Factors such as alpha-element enhancements and varying calibrations of the measurements means that there are large systematic uncertainties when comparing the two metallicity measurements. 
However, given the diffuse nature of the CGM, gradients in the metallicity can only be studied thus far using a combination of emission and absorption. 

\section{Summary and Conclusion}
We have analysed the physical and emission-line properties of 79 galaxies associated with 32 \ion{H}{i} absorbers at redshift $0.20 \lesssim z \lesssim 1.4$. 
These associated galaxies are found at impact parameters ranging from 5.7 to 270 kpc and velocities $-280$ to $+448$ \kms relative to the absorber redshift.
In the sample, there are passive galaxies without detectable emission lines and star-forming galaxies with SFRs up to 15 \Moyr\!.
The emission-line metallicity of associated galaxies also varies from $12 + \log$(O/H) $= 7.62$ to 8.75.
When comparing the properties of associated galaxies with the absorber properties, we find:
\begin{enumerate}
    \item 27 out of the 32 absorbers have galaxy counterparts within $\pm 500$ \kms\!.
    Out of these 27 absorbers, 18 (67 per cent) have two or more associated galaxies and probe overdensities one to three dex larger than the expected number of galaxies within a co-moving volume at $z_{\rm abs}$. 
    However, we note that the non-detections are likely caused by insufficient depth in the MUSE observations for the set of absorbers in the MUSE-ALMA Halos sample and the detection rate is a strict lower limit.
    \item By including results from SINFONI studies of galaxy counterparts to DLA systems, the Bimodal Absorption System Imaging Campaign and the Cosmic Ultraviolet Baryons survey, we find that the \ion{H}{i} column density of absorbers is found to be anti-correlated with the impact parameter $b$ of the nearest associated galaxy out to 120 pkpc. 
    This relationship persists when we scale $b$ by the virial radius of the most probable host galaxy. 
    These results broadly agree with simulations of the circumgalactic medium and we find a $> 3$ dex scatter in column density at any given impact parameter. 
    The scatter highlights the intrinsic `clumpiness' of CGM gas and that absorbers are often not associated with a single galaxy halo. 
    However, the fact that a trend in $N_{\ion{H}{i}}(b)$ can be seen suggests large statistical studies can still reproduce the distribution of gas in the CGM.  
    \item A comparison of the emission metallicity from the ISM of associated galaxies and absorber metal abundance reveals the neutral gas phase in the CGM is lower metallicity than the ISM. 
    From the limited absorber metallicity measurements, there appears to be a negative gradient in the CGM metallicity as we go to larger values of $b$ and $\Delta v$. 
    However, this result is limited by the complications that come with comparing metallicity measurements of two distinct gas phases and elements. 
    \item Adopting a galaxy-centric view and plotting absorbers in velocity difference and impact parameter space indicates that the column density of gas decreases at large values of $b$ and $|\Delta v|$.
    However, we find this is largely driven by the anti-correlation between $N_{\ion{H}{i}}$ and $b$. 
    There appears to be no trend between the \ion{H}{i} column density and the velocity difference between associated galaxy and absorber. 
    This is caused by the uncertainty in the origin of the absorber, that is, whether the neutral gas is found to be co-rotating with the galaxy disk, outflowing or accreting onto the galaxy. 
    Future works will seek to determine the origins of the gas and more deeply relate the absorber properties with the host of the absorber.
\end{enumerate}

The analysis of the 19 MUSE fields from the MUSE-ALMA Halos survey highlight that the increasing number of absorber-galaxy pairs allows for a statistical study of how the gas and metals are distributed in the CGM. 
For the absorbers without associated galaxies, we emphasise the need for deeper observations to improve the completeness of the sample and to follow-up candidate associated galaxies at small $b$ seen in the HST imaging. 
Additionally, higher spatial resolution data of the region near the QSO is essential to uncover passive galaxies that may be responsible for the absorber. 
In future, surveys such as the 4MOST \citep{4MOST, 4MOSTcomm} \ion{H}{i}-Q survey (PI: C. P\'eroux) will seek to increase the number of absorber-galaxy pairs by several orders of magnitude.
The 2.8 Million fiber hour-survey will observe over one million background quasars with spectral resolution $R = 20,000$ in fields with deep observations of foreground objects (galaxies, active galactic nuclei, clusters). Ultimately, these massive statistical samples will allow us to map the gas and metal distribution of the CGM.

\section*{Acknowledgements}
This research is supported by an Australian Government Research Training Program (RTP) Scholarship.
EMS, GK and SW acknowledge the financial support of the Australian Research Council through grant CE170100013 (ASTRO3D).
VPK and AK acknowledge partial support for GO program 15939 (PI: Peroux) provided through a grant from the STScI under NASA contract NAS5-26555 and NASA grant 80NSSC20K0887 (PI: Kulkarni). 
VPK also gratefully acknowledges additional support from the National Science Foundation grants AST/2007538 and  AST/2009811 (PI: Kulkarni). 
We thank the International Space Science Institute (ISSI) (\url{https://www.issibern.ch/}) for financial support.

We thank Thomas Ott for developing and distributing the \textsc{QFitsView} software.
This research also made use of several \textsc{python} packages: \textsc{astropy} \citep{astropy:2013, astropy:2018}, 
\textsc{matplotlib} \citep{Hunter:2007}, \textsc{numpy} \citep{harris2020array} and \textsc{scipy} \citep{2020SciPy-NMeth}.

\section*{Data Availability}
Data directly related to this publication and its figures are available upon request. 
The raw data can be downloaded from the public archives with the respective project codes.



\bibliographystyle{mnras}
\bibliography{MAH_VIII} 
\clearpage


\appendix

\section{MUSE Fields} \label{App:whitelight}
We present MUSE white-light images of the remaining fields in this section. 

\begin{figure*}
    \includegraphics[width=0.77\textwidth]{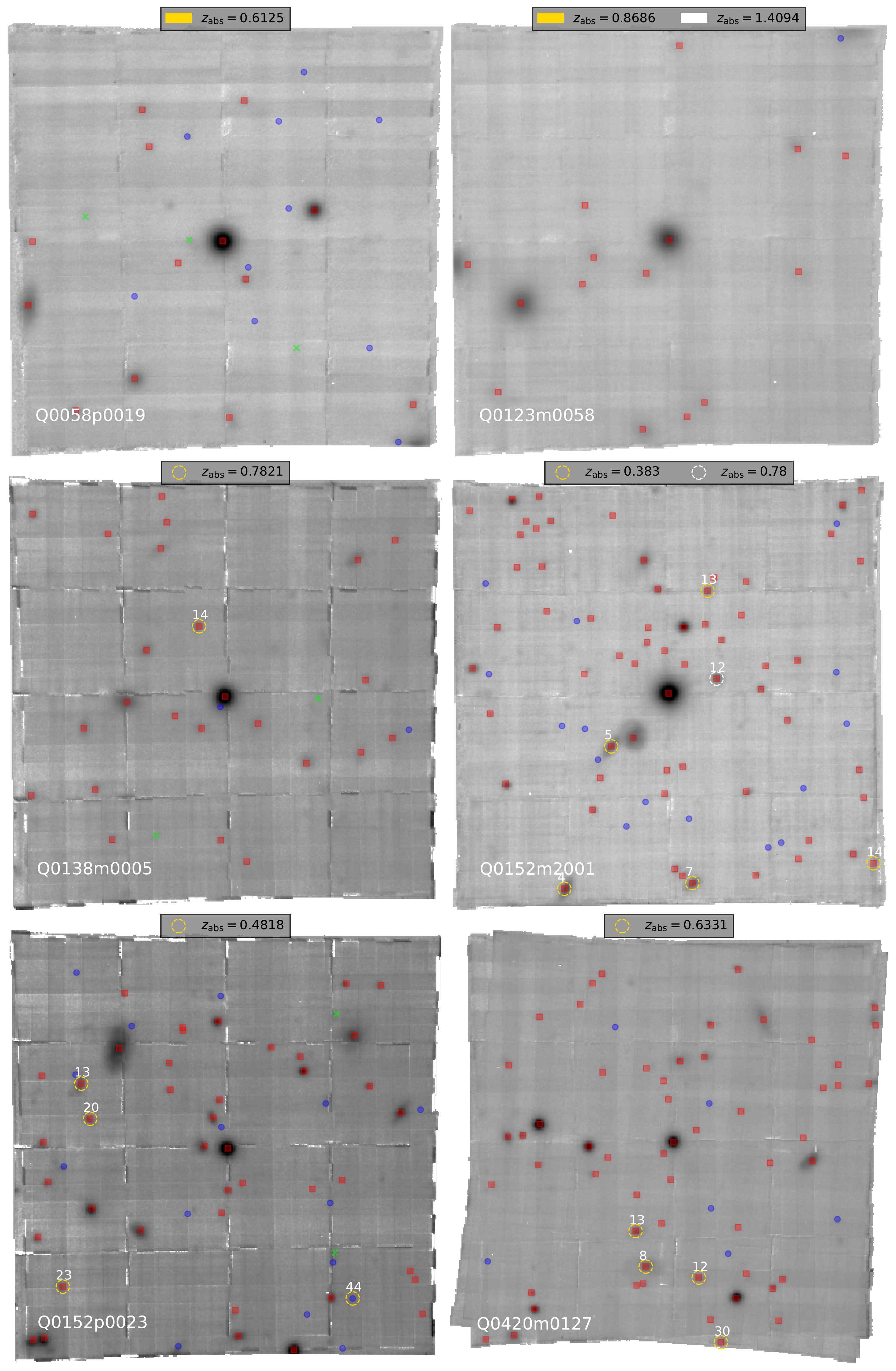}
    \caption{
    White-light images of the MUSE fields centred on the QSO. 
    The four markers represent how the object was detected. 
    Red squares denote continuum objects detected using \textsc{ProFound} and blue circles are emission-line sources. 
    Green crosses are objects found after we used sources in the HST continuum imaging as a prior that were not found initially during our MUSE continuum and emission-line searches. 
    Gold stars represent objects detected using a spectral PSF subtraction of the quasar. 
    We note that many sources will be detected in multiple ways; an emission-line galaxy with bright stellar continuum will be found in MUSE continuum, emission and in the HST continuum imaging. 
    The markers only denote how the source was first detected, following the order of methods listed in \autoref{sec: obs} (using a MUSE continuum and emission-line search, HST continuum detections as a prior and a spectral PSF subtraction of the QSO).
    The dashed circles are galaxies that are within $\pm 500$ \kms of an absorber and they are labelled by their ID in the MUSE-ALMA Halos object catalogues \citep{Peroux2022}.}
    \label{fig:Fields1}
\end{figure*}

\begin{figure*}
    \includegraphics[width=0.8\textwidth]{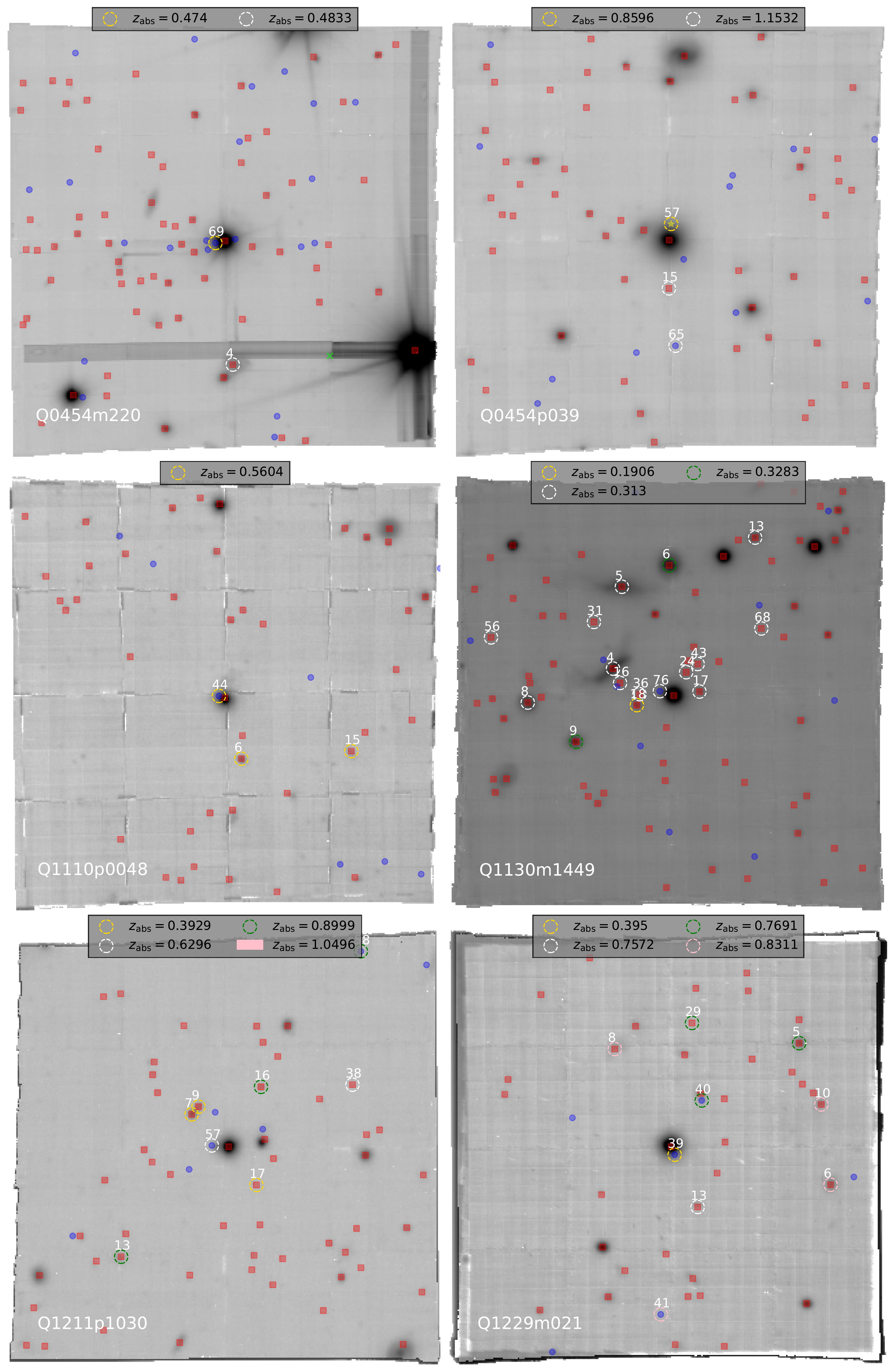}
    \caption{
    See \autoref{fig:Fields1} caption. }
    \label{fig:Fields2}
\end{figure*}

\begin{figure*}
    \includegraphics[width=0.8\textwidth]{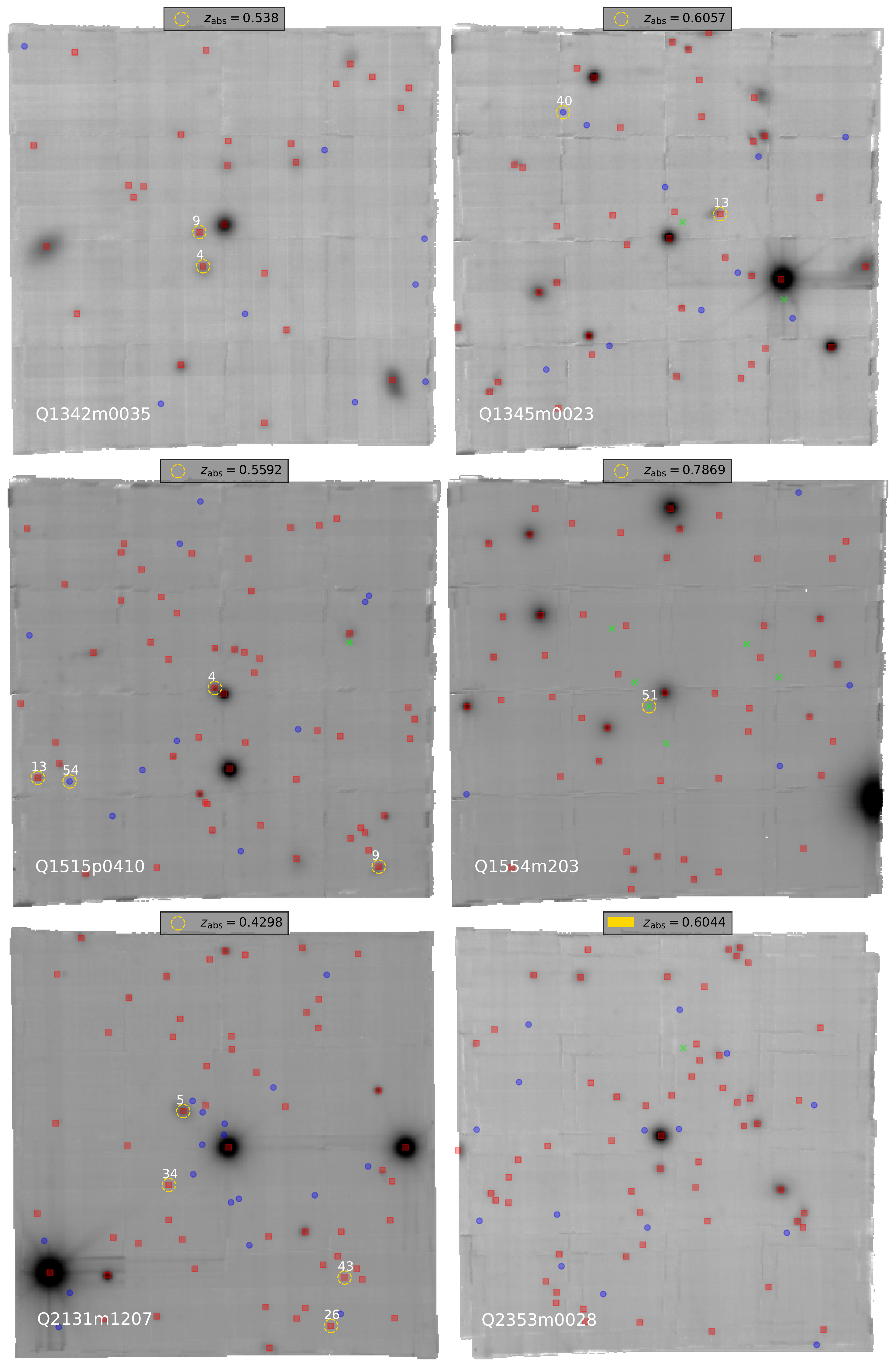}
    \caption{
    See \autoref{fig:Fields1} caption. }
    \label{fig:Fields3}
\end{figure*}

\section{Varying selection criteria for the most probable galaxy host}
\begin{figure*}
    \includegraphics[width=\textwidth]{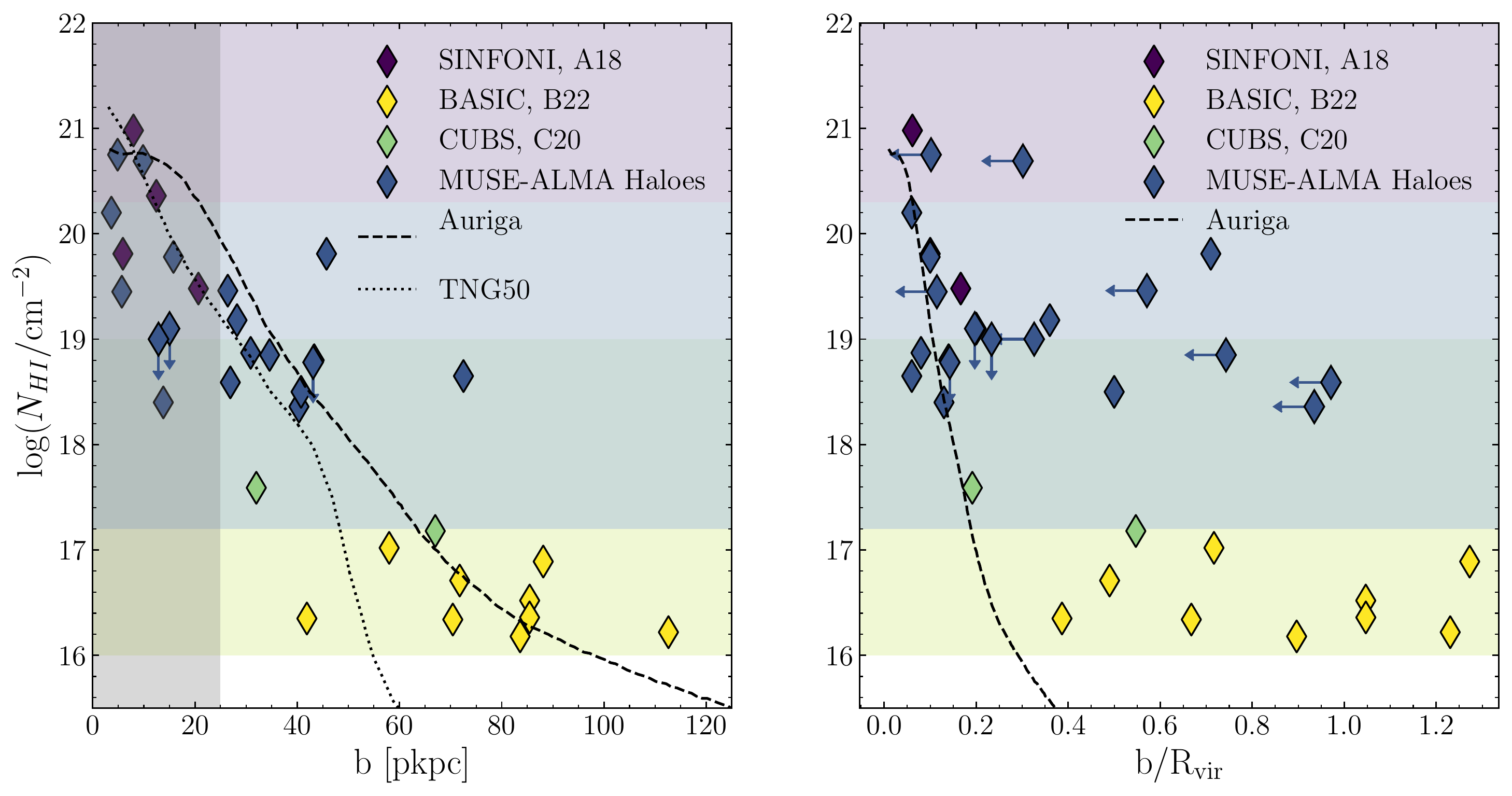}
    \caption{See caption for \autoref{fig:minb_NHI}. 
    We identify the most probable galaxy host of the \ion{H}{i} as the object with the lowest impact parameter and closest in velocity to the absorber. 
    }
    \label{app:minbv_NHI}
\end{figure*}

We check whether varying the criteria used to select the most probable host galaxy of the absorber will change the $N_{\ion{H}{i}}$ versus $b$ relation. 
In \autoref{app:minbv_NHI}, the diamonds represent galaxies that are closest in both impact parameter (minimising $b$) and velocity (minimising $|\Delta v|$) to the absorber. 
18 out of 27 absorbers have an associated galaxy that satisfies both criterion. 
We similarly enforce this criteria for the SINFONI \citep{Augustin2018}, BASIC \citep{Berg2022} and CUBS \citep{Chen2020} surveys. 
In the BASIC survey, the most probable host galaxy was determined by minimizing $\sqrt{(\rho/R_{\rm vir})^2 + (|\Delta v|/v_{\rm esc})^2}$, where $R_{\rm vir}$ is the virial radius and $v_{\rm esc}$ the escape velocity of the galaxy halo \citep{Berg2022}. 
The associated galaxies selected by minimising $b$ and $\Delta v$ are found to be a sub-sample of the galaxies from the BASIC survey. 

\section{Changing \ion{H}{i} column density with impact parameter and velocity difference to the absorber}
\label{App:calc}
\begin{figure*}
    \includegraphics[width=\textwidth]{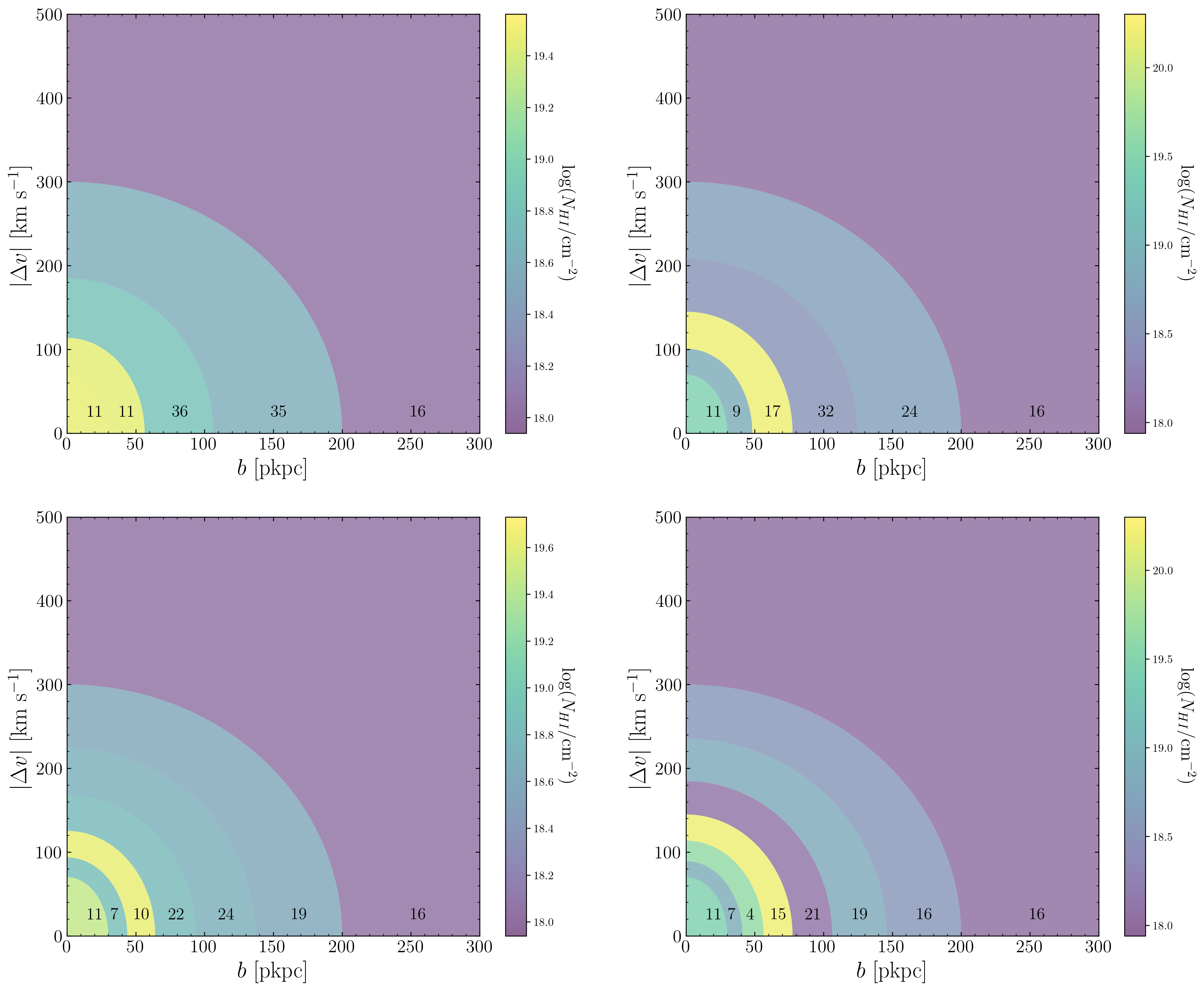}
    \caption{The median $\logNHI$ column density in two-dimensional bins of velocity difference to the absorber redshift and impact parameter. 
    We try splitting the parameter space into five to eight radial bins. 
    The innermost bin is kept constant and covers up to $\Delta v = 70$ \kms and $\rho = 30$ pkpc. 
    The remaining bin sizes are distributed uniformly in log space, and the number count in each bin is given by the value above the $x$-axis. 
    In this plot, we include all 113 galaxies associated with \ion{H}{i} absorbers from the MUSE-ALMA Halos \citep{Peroux2022}, CUBS \citep{Chen2020} and BASIC \citep{Berg2022} surveys.}
    \label{fig:HI_wQ1130}
\end{figure*}

\begin{figure*}
    \includegraphics[width=\textwidth]{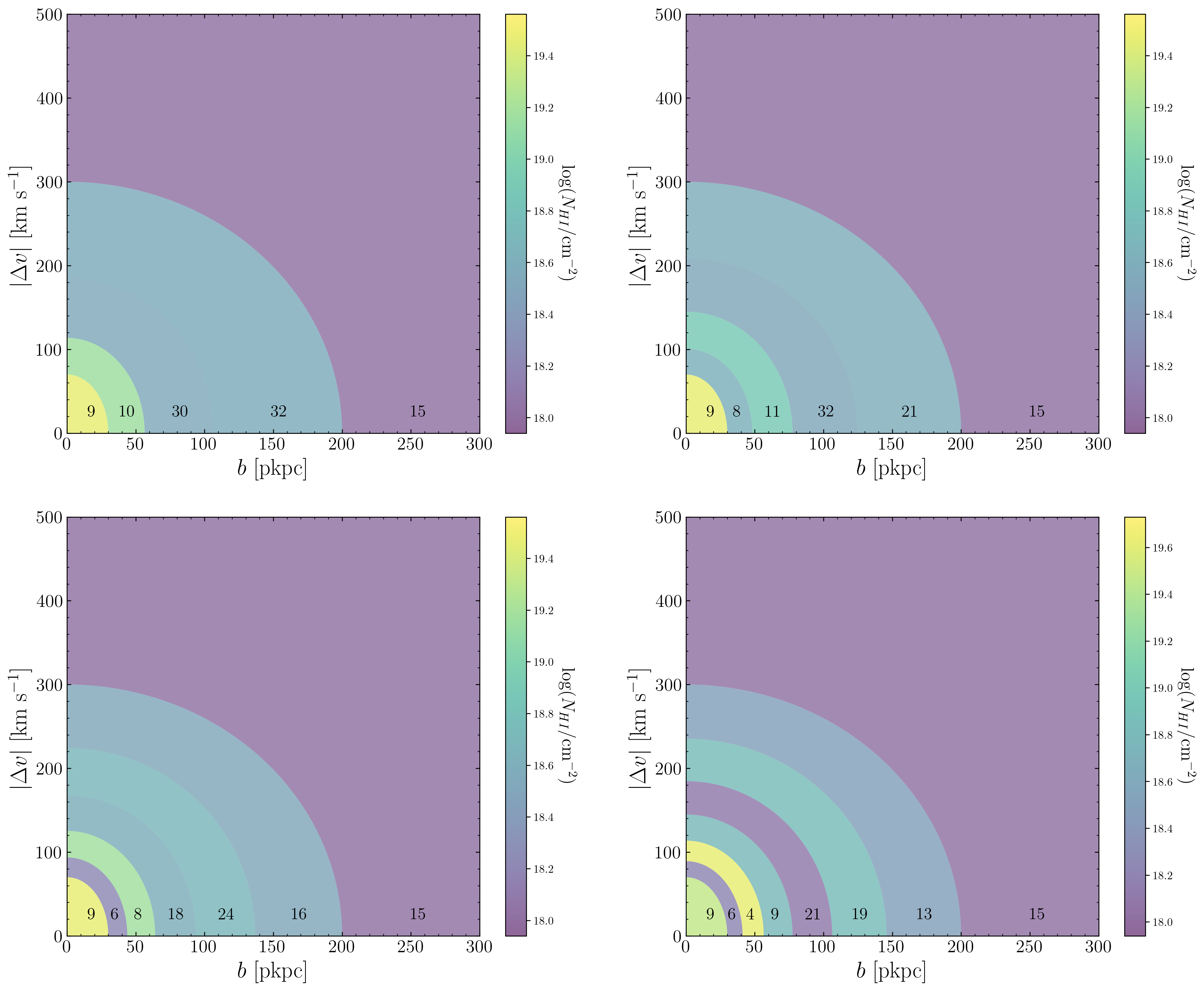}
    \caption{
    We remove the 13 galaxies associated with the $z_{\rm abs} = 0.313$ and $\logNHI = 21.7$ cm$^{-2}$ absorber towards Q$1130-1449$ and replicate the plot in \autoref{fig:HI_wQ1130}. 
    This is done to check whether the trend of declining column density with larger impact parameters and velocity differences continues to exist without the inclusion of this high column density system.}
    \label{fig:HI_noQ1130}
\end{figure*}

We vary the bins used in the top-right plot of \autoref{fig:densities} to test whether the decrease in \ion{H}{i} column density towards larger $b$ and $|\Delta v|$ is artificially caused by the binning.
In \autoref{fig:HI_wQ1130}, we use an inner bin covering up to $|\Delta v| = 70$ \kms and $\rho = 30$ pkpc before dividing the remaining parameter space uniformly into four to seven bins.
The trend appears to persist and is most evident when five or six bins are used. 
We still find that $N_{\ion{H}{i}}$ at larger $\rho$ and $|\Delta v|$ is lower despite the variations in the column density within the inner $\sim$70 pkpc when more bins are used.
This exercise is repeated in \autoref{fig:HI_noQ1130} where we exclude the $13$ galaxies associated with the absorber towards Q1130$-$1449 with \logNHIunit $= 21.71$. 
While the median column densities at lower impact parameters are smaller, the trend still persists.

\begin{figure*}
    \includegraphics[width=\textwidth]{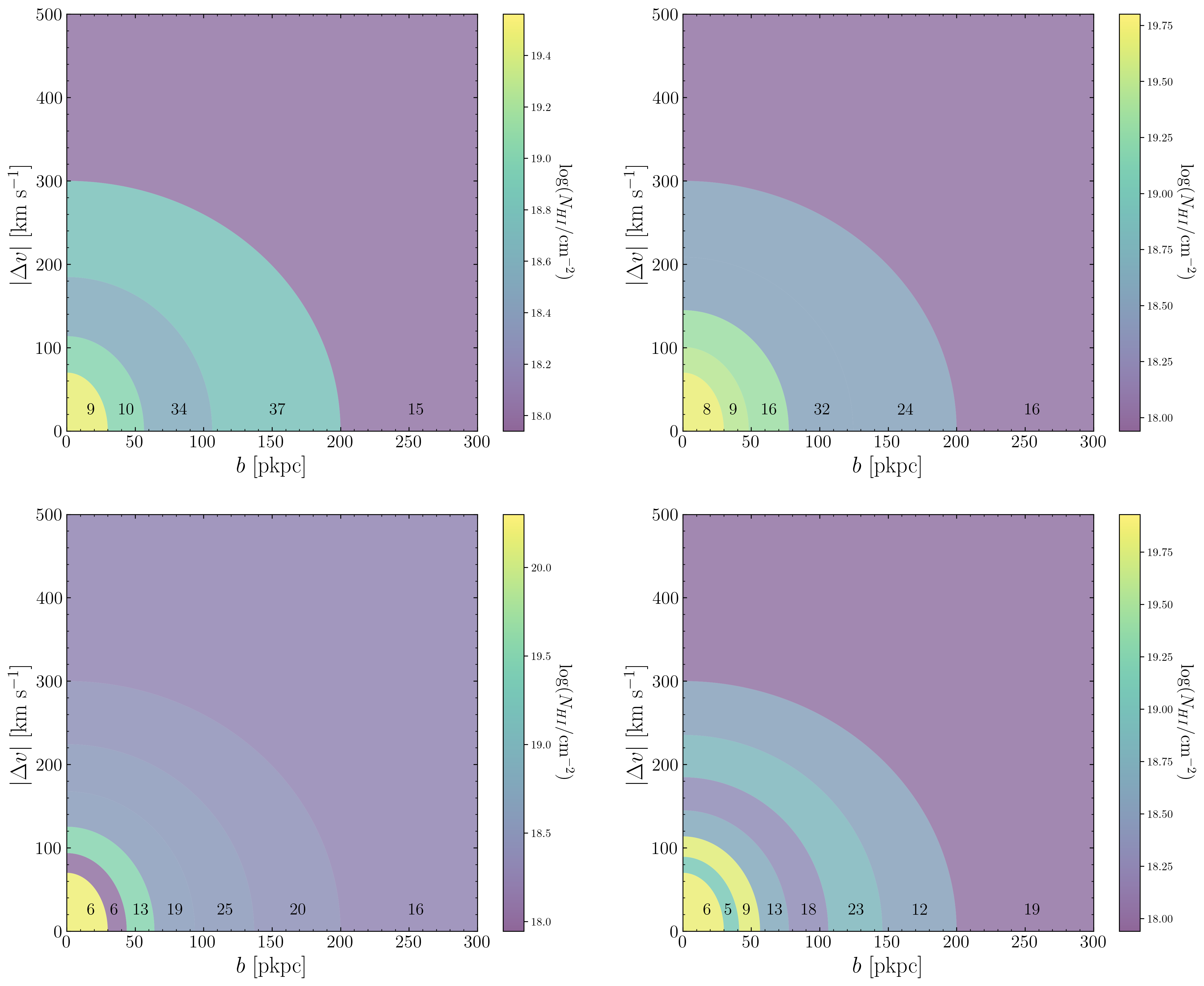}
    \caption{
    To test the contribution of $\Delta v$ to the declining \ion{H}{i} column density at larger bins, we randomly scramble our velocity difference values and replicate the plot in \autoref{fig:HI_wQ1130}. 
    We find that $\Delta v$ does not affect the decreasing trend of \ion{H}{i} significantly. 
    }
    \label{fig:HI_randomV}
\end{figure*}

\begin{figure*}
    \includegraphics[width=\textwidth]{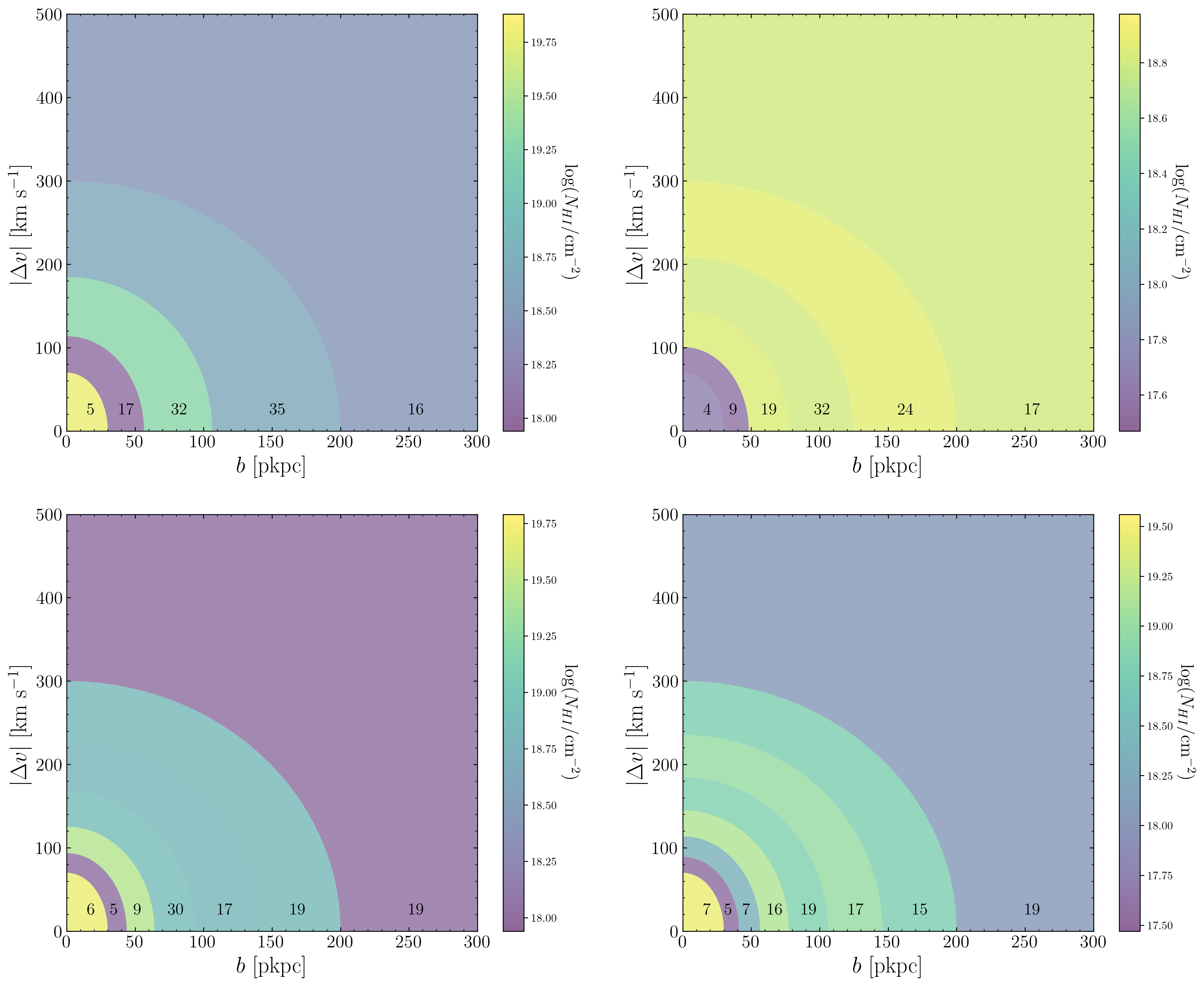}
    \caption{
    To test the contribution of $\rho$ to the declining \ion{H}{i} column density at larger bins, we randomly scramble our impact parameter values and replicate the plot in \autoref{fig:HI_wQ1130}. 
    The lack of trend seen indicates the impact parameter plays a larger role when measuring the distribution of gas in the CGM. 
    }
    \label{fig:HI_randomb}
\end{figure*}

From \autoref{fig:minb_NHI} and \autoref{fig:dv_NHI}, we find that while $N_{\ion{H}{i}}$ decreases with impact parameter, there is no such trend in the $\logNHI$ versus $|\Delta v|$ plot.  
Here, we test whether randomly scrambling our values of $|\Delta v|$ and $b$ will affect the distributions seen in the top-right plot of \autoref{fig:densities}. 
The plots in \autoref{fig:HI_randomV} and \autoref{fig:HI_randomb} confirm that the impact parameter plays a larger role when measuring the distribution of gas in the CGM. 
Scrambling our velocity difference values does not affect the trends we see, whereas the decrease in \ion{H}{i} column density at larger $b$ and $|\Delta v|$ can no longer be seen in \autoref{fig:HI_randomb} after shuffling the impact parameter values.

Finally, we normalise our velocity difference and impact parameter values to the escape velocity ($v_{\rm esc}$) and virial radius $R_{\rm vir}$ respectively. 
$v_{\rm esc}$ is defined as the velocity required to escape the halo from the radius of maximum circular velocity \citep[Equation 16 in][]{Shull2014} assuming a Navarro-Frenk-White (NFW) profile \citep{NFW1997}. 
To calculate the virial radius, we first convert the stellar mass to a virial mass using the stellar-to-halo mass relation in \citet{Puebla2017}. 
From here, we adopt $R_{\rm vir}$ as the radius where the enclosed density is 200 times the critical density of the Universe at the redshift of the associated galaxy ($R_{200}$). 
We find that the trend of declining \ion{H}{i} column density with increasing $|\Delta v|/v_{\rm esc}$ and $b/R_{\rm vir}$ persists for values of $|\Delta v|/v_{\rm esc} < 1$ and $b/R_{\rm vir} < 3$. 
This is understandable because beyond these normalised values, the gas is likely no longer associated with a galaxy's CGM. 
Additionally, our combined sample is limited to \ion{H}{i} column densities $\logNHIunit > 16.0$; we might expect the trend to continue if lower column densities are probed. 

\begin{figure*}
    \includegraphics[width=\textwidth]{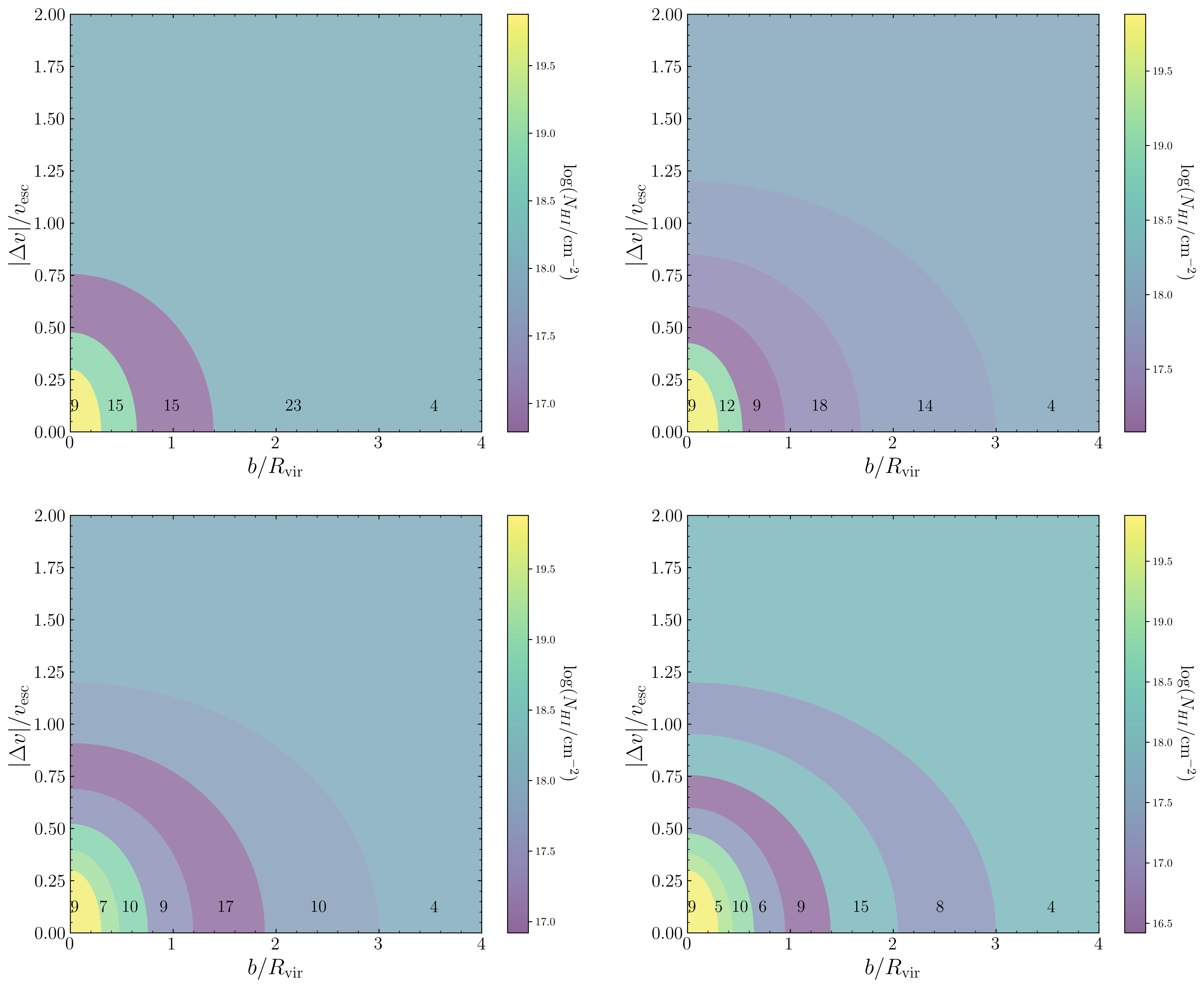}
    \caption{
    For all galaxies with stellar masses, we compute the virial radius $R_{\rm vir} = R_{\rm 200}$ and  escape velocity assuming a NFW profile. 
    Then, we scale our $\Delta v$ and $\rho$ values and replicate \autoref{fig:HI_wQ1130}.
    The trend of declining \ion{H}{i} column density with increasing $|\Delta v|/v_{\rm esc}$ and $b/R_{\rm vir}$ persists for values of $|\Delta v|/v_{\rm esc} < 1$ and $b/R_{\rm vir} < 3$. 
    Beyond these normalised values, the gas is likely no longer associated with a galaxy's CGM. 
}
    \label{fig:HI_norm}
\end{figure*}


\bsp	
\label{lastpage}
\end{document}